\documentclass[11pt,onecolumn]{article}
\usepackage{microtype}
\usepackage{graphicx}
\usepackage{subfigure}
\usepackage{multirow}
\usepackage{threeparttable}
\usepackage{float}
\usepackage{booktabs} 

\usepackage{hyperref}

\usepackage{amsmath}
\usepackage{amssymb}
\usepackage{mathtools}
\usepackage{amsthm}

\def \bX {{\mathbf{X}}}
\def \bx {{\mathbf{x}}}

\def \bz {{\mathbf{z}}}

\def \bC {{\mathbf{C}}}

\def \bH {{\mathbf{H}}}

\def \bu {{\mathbf{u}}}
\def \bp {{\mathbf{p}}}

\def \cH {{\mathcal{H}}}

\newcommand\blfootnote[1]{%
  \begingroup
  \renewcommand\thefootnote{}\footnote{#1}%
  \addtocounter{footnote}{-1}%
  \endgroup
}
\usepackage[capitalize,noabbrev]{cleveref}

\theoremstyle{plain}
\newtheorem{theorem}{Theorem}[section]
\newtheorem{proposition}[theorem]{Proposition}
\newtheorem{lemma}[theorem]{Lemma}
\newtheorem{corollary}[theorem]{Corollary}
\theoremstyle{definition}
\newtheorem{definition}[theorem]{Definition}
\newtheorem{assumption}[theorem]{Assumption}
\theoremstyle{remark}
\newtheorem{remark}[theorem]{Remark}

\usepackage[textsize=tiny]{todonotes}

\usepackage{authblk}

\title{Towards a Foundation Model for Brain Age Prediction using coVariance Neural Networks}

\begin{document}

\author[1]{ Saurabh Sihag}
\author[2]{ Gonzalo Mateos}
\author[1]{ Alejandro Ribeiro}

\affil[1]{\small University of Pennsylvania, Philadelphia, PA.}
\affil[2]{\small University of Rochester, Rochester, NY.}
\date{}
\maketitle




\begin{abstract}
Brain age is the estimate of biological age derived from neuroimaging datasets using machine learning algorithms. Increasing brain age with respect to chronological age can reflect increased vulnerability to neurodegeneration and cognitive decline. In this paper, we study NeuroVNN, based on coVariance neural networks, as a paradigm for foundation model for the brain age prediction application. NeuroVNN is pre-trained as a regression model on healthy population to predict chronological age using cortical thickness features and fine-tuned to estimate brain age in different neurological contexts. Importantly, NeuroVNN adds anatomical interpretability to brain age and has a `scale-free' characteristic that allows its transference to datasets curated according to any arbitrary brain atlas. Our results demonstrate that NeuroVNN can extract biologically plausible brain age estimates in different populations, as well as transfer successfully to datasets of dimensionalities distinct from that for the dataset used to train NeuroVNN. 
\end{abstract}
\blfootnote{Preliminary work. All datasets used in this paper are publicly available. Access to them can be obtained through their respective data access protocols. Contact \href{sihag.saurabh@gmail.com}{sihag.saurabh@gmail.com} for the NeuroVNN model and code used for results reported in this manuscript.}
\section{Introduction}
Aging is a convoluted biological process that manifests in the form of various progressive physiological and cognitive changes~\cite{lopez2013hallmarks}. Importantly, ageing is associated with increased likelihood of morbidity and mortality due to decline in structural, functional, and cognitive integrity~\cite{fjell2010structural,cole2018brain,world2015world}, thus, leading to elevated healthcare-related and socio-economic burden for a society with an ageing population~\cite{world2015world}. To address such challenges, there has been an increasing focus on characterizing the transition from healthy stage to disease onset at the individual level for various age-related health conditions through studies on biomarkers~\cite{johnson2009longitudinal,ferrucci2020measuring}. In this context, brain-predicted biological age, also referred to as \emph{brain age}, has emerged as a biomarker of interest that can characterize the biological age for an individual based on the assessment of their neuroimaging data~\cite{cole2017predicting,cole2018brain,baecker2021machine,baecker2021brain,beheshti2021predicting,pina2022structural, franke2019ten, franke2012longitudinal,sihag2023explainable}.

Recent years have seen an exponential rise in the study of brain ageing using machine learning algorithms~\cite{baecker2021machine}. A common objective of brain age prediction strategies is to derive an estimate for brain age from neuroimaging data and compare it with the chronological age (time since birth) via \emph{brain age gap}, i.e., the difference between brain age and chronological age. A large brain age gap has been demonstrated to be indicative of accelerated aging in various neurological diseases, thus, implying higher disease burden and risk of mortality~\cite{cole2018brain}. Subsequently, we use the notation $\Delta$-Age to refer to brain age gap.  

\begin{figure}[t!]
  \centering
  \includegraphics[scale=0.25]{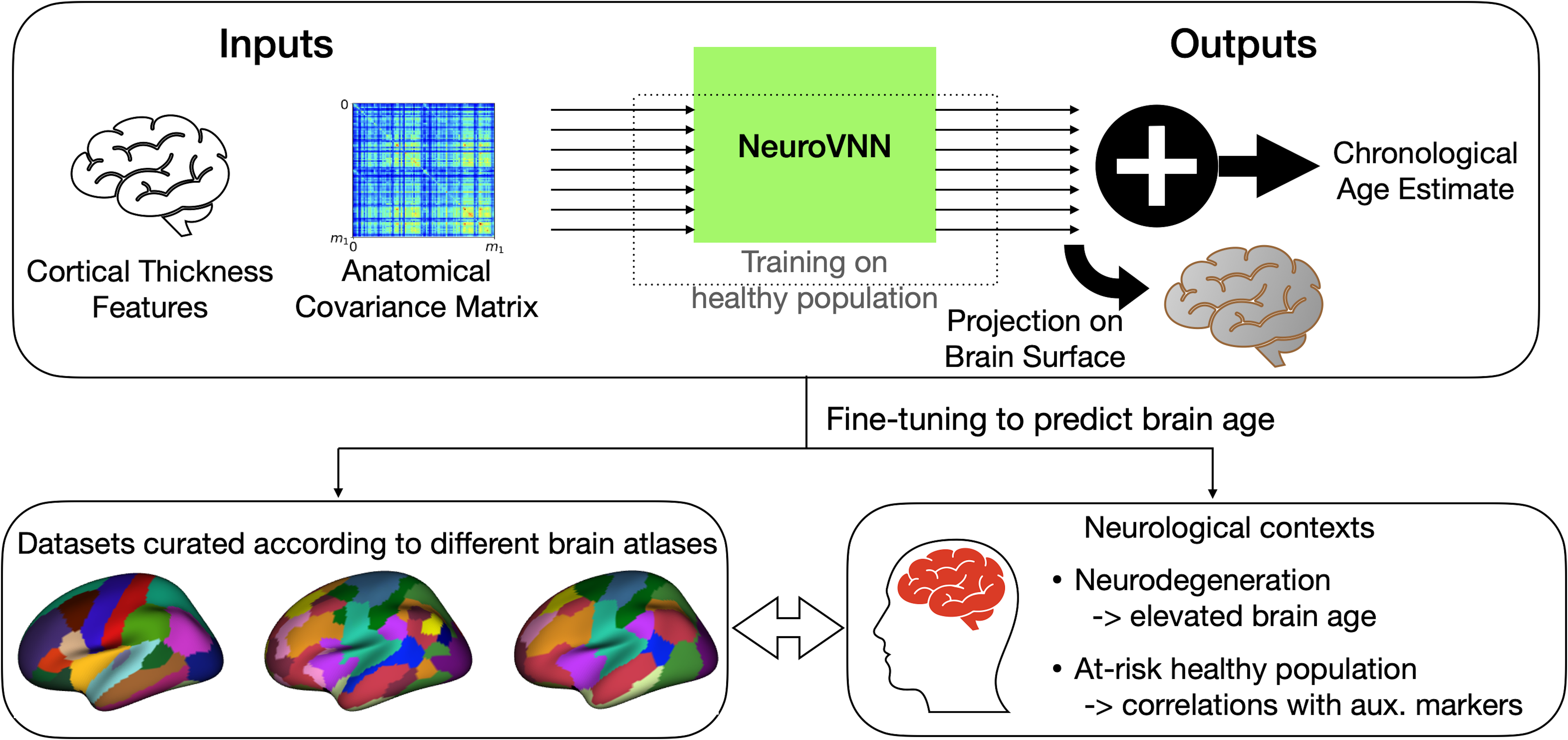}
   \caption{An overview of NeuroVNN as the paradigm for foundation model for brain age prediction.}
   \label{brain_age_vnn}
\end{figure}

\noindent
{\bf Brain age prediction algorithms and foundation models.} In principle, the prevalent algorithmic frameworks for brain age prediction draw similarities with the `pre-training then fine-tuning' characteristic of foundation models~\cite{bommasani2021opportunities}. Specifically, the core of a brain age prediction algorithm is a regression model that is pre-trained to predict chronological age for a healthy population by leveraging their multivariate neuroimaging data. When deployed on a wider population with appropriate fine-tuning, it is expected that the performance of this regression model will degrade in a structured manner in various neurological contexts (such as disease, genetic disposition, psychiatric factors, etc.) and yield different statistical patterns in $\Delta$-Age~\cite{lee2022deep,karim2021aging,liem2017predicting,ronan2016obesity, mareckova2020maternal,westlye2012effects}. 

\noindent
{\bf How can foundation models contribute to brain age prediction?}
Brain age prediction algorithms are likely to provide the most significant impact in clinical studies that track different biological variables, including neuroimaging, in a focused population to address specific biological questions (for instance, tracking an at-risk healthy population due to their genetic disposition). Due to high cost of data acquisition, the number of participants or data samples in these studies is limited and often, not enough to train credible machine learning models~\cite{thomas2020dealing}. Therefore, a foundation model trained on a representative healthy population can provide a feasible way to leverage the representative power of sophisticated machine learning algorithms to derive $\Delta$-Age on datasets of limited size and unify the findings reported across different contexts.

The prevalence of different brain atlases or maps used to curate neuroimaging datasets is another limiting factor on the sample size and applicability of machine learning models in data analyses~\cite{lawrence2021standardizing}. Specifically, each brain atlas provides a unique map for brain organization and hence, can have a different number of features. Although these brain atlases can provide unique insights into brain organization, they also create challenges in data analyses due to limitations imposed on the transferability of machine learning models to datasets of different dimensionalities~\cite{yang2022data}. Thus, a foundation model that can effectively transfer its capabilities of inferring meaningful $\Delta$-Age to datasets with dimensionalities different from that for its training dataset can help unify the analyses across different brain atlases, improve reproducibility of findings, and furnish the understanding of biomarkers, such as $\Delta$-Age, from different perspectives simultaneously.

\noindent
{\bf coVariance Neural Networks as foundation models.} 
coVariance neural networks (VNN) have been studied as graph convolutional networks that operate on sample covariance matrices from the data as the graph~\cite{sihag2022covariance}. Pertinent to this paper, we make three observations from the existing work on VNNs: (i) the VNN models can be structured as \emph{scale-free} regression models that can process dataset of any arbitrary dimensionality~\cite{sihag2023transferablility}; (ii) the architecture of VNNs has been leveraged to provide anatomically interpretable brain age estimates in AD~\cite{sihag2023explainable}, where the interpretability offered was shown to be an in-built feature of VNNs; and (iii) the inference outcomes by VNNs exhibit robustness to the number of samples used to train the model~\cite{sihag2022covariance}. The interpretability offered by VNNs is in contrast to other data-specific approaches (such as SHAP, LIME, saliency maps) studied for explainability of deep learning models in the literature for brain age prediction application~\cite{lombardi2021explainable,yin2023anatomically,lee2022deep}. The robustness offered by VNN is pertinent as the number of data samples in biomedical applications is usually limited and hence, it is critical to have foundation models whose inference outcomes are robust to variations in the size of the datasets and reproducible. 

Because VNNs can be set up as scale-free regression models, it is practically feasible to investigate the generalizability of their findings from the training dataset to another dataset of any arbitrary dimensionality. Thus, VNNs provide a unique ability to unify the analyses across datasets that capture similar information about the brain at different resolutions and using distinct maps.

%


\noindent
{\bf Contributions.} In this paper, we propose a coVariance Neural Network (VNN)-based paradigm for a foundation model, termed as NeuroVNN, for estimating brain age. NeuroVNN is pre-trained to predict chronological age for a healthy population spanning the adult life span ($18$ years and older) using the cortical thickness features. After pre-training, NeuroVNN can be transferred to datasets specific to various neurological contexts and those curated according to different brain atlases. Specifically, our experiments demonstrate that (i) the outputs of NeuroVNN can be fine-tuned to derive biologically meaningful $\Delta$-Age estimates across healthy populations at-risk of developing dementia (due to family history or amyloid status) as well as in a neurodegenerative context; and (ii) NeuroVNN provides anatomical profiles associated with the observed statistical patterns in $\Delta$-Age estimates and these anatomical profiles can be recovered across datasets curated according to various brain atlases via one-shot transferability. Thus, the observations (i) and (ii) from our experiments highlight the capabilities of NeuroVNN as a viable foundation model for this application. Further, we also explore the practical considerations in deploying NeuroVNN on datasets that may be skewed towards an age group that is relatively sparsely represented in the training set used to pre-train NeuroVNN. Together, the results in this paper demonstrate that NeuroVNN can provide a baseline to construct foundation models for biomarker discovery, that can be built upon in future works in computational neuroscience.

\section{Related Work}\label{lit_review}
Here, we provide a review of relevant existing literature and positioning of this paper relative to it.

\noindent
{\bf Foundation models.} Foundation models have provided a significant shift in practical deployment of artificial intelligence methods; enabling a shift from context-specific models with narrow applicability to models that can readily be adapted to downstream tasks spanning versatile applications or contexts~\cite{brown2020language, yuan2021florence, yu2022coca,wu2023visual,zhou2023comprehensive,kirillov2023segment}. Broadly, foundation model has been used as an umbrella term for machine learning models that are pre-trained on a large dataset and their learned parameters held fixed, while they are applied to downstream tasks across various domains or tasks; usually with characteristics unseen during training. 

Foundation models are effective in applications where the downstream tasks are limited by sample size constraints and, hence, attractive to deployment as a machine learning framework for applications in healthcare~\cite{moor2023foundation, wojcik2022foundation}. A few recent efforts have been made in developing language-based foundation models for electronic health records~\cite{steinberg2021language,yang2022data}. In this paper, we propose NeuroVNN as a foundation model for brain age prediction application that is effective in addressing the heterogeneity encountered in neuroimaging datasets and neurological contexts. In particular, NeuroVNN is based on the scale-free VNN model which can transfer readily across datasets curated according to different brain atlases without change in its architecture. To the best of our knowledge, there is no comparable machine learning model that can seamlessly generalize to datasets of different dimensionalities in this context.  

\noindent
{\bf coVariance Neural Networks.} VNNs exploit the eigenspectrum of the sample covariance matrix to achieve learning objective~\cite{sihag2022covariance}. Importantly, VNNs are provably stable to the number of samples used to estimate the covariance matrix, hence, adding confidence to the reproducibility of their findings across datasets of different sizes. Moreover, theoretical properties of the transferability of VNNs across datasets of different dimensionalities has been studied in~\cite{sihag2023transferablility}. VNNs have been demonstrated to successfully transfer across multi-resolution neuroimaging datasets in~\cite{sihag2022covariance} and~\cite{sihag2022predicting}. Recently, a VNN-based brain age prediction framework has been proposed in~\cite{sihag2023explainable}, which adds an anatomically interpretable and methodologically explainable perspective to the brain age prediction task. 

The scope of results in this paper extends significantly beyond that of existing studies on VNNs in various aspects. Firstly, NeuroVNN was trained on a more holistic dataset that spanned a broader adult life span (18 years and older), unlike the previous studies that have focused exclusively on older adults ($50$ years or older). Furthermore, we demonstrate the utility of brain age estimated using NeuroVNN on datasets with healthy population characterized by elevated risk of dementia. Previous works on brain age prediction using VNNs have focused solely on detecting elevated brain age in AD. Finally, our experiments reveal the practical challenges and potential solutions to them when NeuroVNN was deployed on datasets concentrated in age groups that were relatively under-represented in the training set used to train NeuroVNN. Together, these aspects along with the anatomical interpretability offered by VNNs on brain age prediction task provide a more holistic study of the applicability of NeuroVNN in the brain age prediction application.

\noindent
{\bf Brain age prediction.} There exist numerous machine learning studies for brain age prediction using neuroimaging data. We refer the reader to~\cite{more2023brain} for a detailed overview. Brain age, by itself, does provide diagnostic specificity as brain age gap can be elevated due to a variety of factors in addition to diseases. The VNN-based framework in~\cite{sihag2023explainable} addressed this concern, as its architecture could be set up to provide in-built anatomical interpretability to the brain age estimate. Without anatomic specificity of $\Delta$-Age, concerns may arise pertaining to the role of age-bias correction step~\cite{butler2021pitfalls} or variations induced in brain age due to various factors~\cite{more2023brain}. Such concerns are more profound in brain age prediction algorithms deployed as a black-box, with little to no clarity in what factors led to elevated brain age gap. The anatomical interpretability offered by a VNN-based architecture can help cross-validate or verify the biological plausibility of inferred brain age gap.  

\section{coVariance Neural Networks}\label{vnn_intro}
We first provide a brief description of VNN models. The architecture of VNNs is inherited from graph convolutional networks~\cite{gama2020graphs} and their convolution operation is characterized by the sample covariance matrix estimated from the multivariate dataset~\cite{sihag2022covariance}. A dataset $\bX_n = [\bx_1,\dots,\bx_n]$ consisting of $n$ independent and identically distributed (i.i.d) $m$-dimensional samples can be leveraged to estimate the sample covariance matrix $\bC$ of size $m\times m$ as $\bC \triangleq \frac{1}{n-1} \sum\limits_{i=1}^n(\bx_i - \bar\bx) (\bx_i-\bar\bx)^{\sf T}$, where $\bar\bx$ is the sample mean of samples in $\bX_n$. 

\noindent
{\bf Convolution using coVariance filters.} The convolution operation in a VNN is modeled by a coVariance filter, given by $\bH(\bC) \triangleq \sum_{k=0}^K h_k \bC^k$, where the scalar parameters $\{h_k\}_{k=0}^K$ are the \emph{filter taps} that are learned from the data during training. 
For $K>1$, the convolution operation leverages filter taps $\{h_k\}_{k=0}^K$ to combine information in a multivariate sample $\bx \in {\mathbb R}^{m\times 1}$ across multi-hop neighborhoods (up to $K$-hop; determined according to covariance matrix $\bC$) to form the output $\bz = \bH(\bC)\bx$.



A single layer of VNN is formed by concatenating the coVariance filter with a point-wise non-linear activation function $\sigma(\cdot)$ (e.g., ${\sf ReLU}, \tanh$) that satisfies $\sigma(\bu) = [\sigma(u_1), \dots, \sigma(u_m)]$ for $\bu = [u_1, 
 \dots, u_m]$. Thus, the output of a single layer VNN for an input $\bx$ is $\bz = \sigma(\bH(\bC) \bx)$. A simple multi-layer VNN architecture can be formed by concatenating multiple one-layer VNNs in series.

\noindent
{\bf Multi-layer VNN.} For a VNN with $L$-layers, denote the coVariance filter in layer $\ell$ by~$\bH_{\ell}(\bC)$ and its corresponding set of filter taps by $\cH_{\ell}$. Given a pointwise nonlinear activation function $\sigma(\cdot)$, the relationship between the input $\bx_{\ell-1}$ and the output $\bx_{\ell}$ for the $\ell$-th layer is
    \begin{align}
        \bx_{\ell} = \sigma(\bH_{\ell}(\bC)\bx_{\ell-1})\,\quad\text{for}\quad \ell\in \{1,\dots,L\}\;,
    \end{align} 
    where $\bx_0$ is the input $\bx$. 
Furthermore, the expressive power of VNN can be increased by incorporating multiple input multiple output (MIMO) processing at every layer. To formalize MIMO processing in VNN, we consider a VNN layer $\ell$ that consists of $F_{\ell-1} \times F_{\ell}$ number of filter banks and is set up to process $F_{\ell-1}$ number of $m$-dimensional inputs to yield $F_{\ell}$ number of $m$-dimensional outputs~\cite{gama2020stability}. In this setting, if the input is specified as $\bX_{\sf in} = [\bx_{\sf in}[1],\dots,\bx_{\sf in}[F_{\sf in}]]$, and the output is specified as $\bX_{\sf out} = [\bx_{\sf out}[1],\dots,\bx_{\sf out}[F_{\sf out}]]$, the relationship between the $f$-th output $\bx_{\sf out}[f]$ and the input $\bx_{\sf in}$ is governed by $\bx_{\sf out}[f] =  \sigma\Big(\sum_{g = 1}^{F_{\sf in} } \bH_{fg}(\bC)\bx_{\sf in} [g] \Big)$,
where $\bH_{fg}(\bC)$ is the coVariance filter that processes $\bx_{\sf in}[g]$. Without loss of generality, we assume that $F_{\ell} = F,\forall \ell \in \{1,\dots,L\}$. 

We use the notation $\Phi(\bx;\bC,{\cal H})$ to compactly represent a VNN, where the set of filter taps $\cH$ captures the full span of its architecture. We also use the notation $\Phi(\bx;\bC,{\cal H})$ to denote the output at the final layer of the VNN. The VNN final layer output $\Phi(\bx;\bC,{\cal H})$ is succeeded by a readout function that maps it to the desired inference outcome.


\noindent
{\bf VNN as a scale-free regression model.} In this paper, the VNN model is deployed as a regression model, where $m$ cortical thickness features are leveraged to estimate chronological age. Due to MIMO functionality, the VNN architecture has $F$ number of $m$-dimensional outputs in the final layer, i.e., $\Phi(\bx;\bC,\cH)$ is of dimensionality $m \times F$. The regression output is determined by a readout layer, which evaluates an unweighted mean of all outputs at the final layer of VNN. Therefore, the estimate $\hat y$ for input $\bx$ is given by
\begin{align*}
    \hat y = \frac{1}{m}\sum\limits_{j = 1}^m \bp_j\;\quad\text{where}\quad \bp_j =  \frac{1}{F} \sum\limits_{f=1}^F [\Phi(\bx;\bC,\cH)]_{jf}\;,
\end{align*}
where $[\Phi(\bx;\bC,\cH)]_{jf}$ denotes the $j$-th output at the $f$-th filter bank in $\Phi(\bx;\bC,\cH)$. The equation above implies that the estimate $\hat y$ can be perceived as the unweighted mean of an $m$-dimensional vector, whose elements $[\bp_j]_{j=1}^m$ are determined by the mean of the outputs of all filters in the filter bank at the final layer. In the experiments performed in this paper, the multivariate input $\bx$ correspond to cortical thickness features for an individual, with each element of $\bx$ associated with a unique cortical region. The output $\hat y$ forms the estimate for chronological age for an individual with cortical thickness features $\bx$.  Since there is no change in dimensionality from the input to the final layer of VNN, projecting the elements $\{\bp_j\}$ on the respective brain regions associated with their indices at the input can provide an anatomical perspective in terms of individual contributions made by brain regions to estimate $\hat y$. 

In principle, the filter taps $\cH$ are independent of the dimensionality $m$ and hence, the VNN $\Phi(\bx;\bC,\cH)$ can process a dataset of any arbitrary dimensionality $m$. Furthermore, the outputs of VNN are expected to be robust to the number of samples used to estimate the sample covariance matrix $\bC$~\cite{sihag2022covariance}.


\section{NeuroVNN as Foundation Model for Brain Age Prediction}
We propose NeuroVNN as a foundation model paradigm for brain age prediction application that can generalize across different neurological contexts and datasets of different dimensionalities. NeuroVNN is a regression model pre-trained to predict chronological age of healthy population. In this paper, we focus on cortical thickness data derived from structural MRI to train NeuroVNN. In principle, any multivariate dataset representing brain morphometry (such as area, volume, etc.) could be adopted in the framework presented here. 
\subsection{Training Datasets and Preprocessing}
\begin{figure}[h]
  \centering
  \includegraphics[scale=0.4]{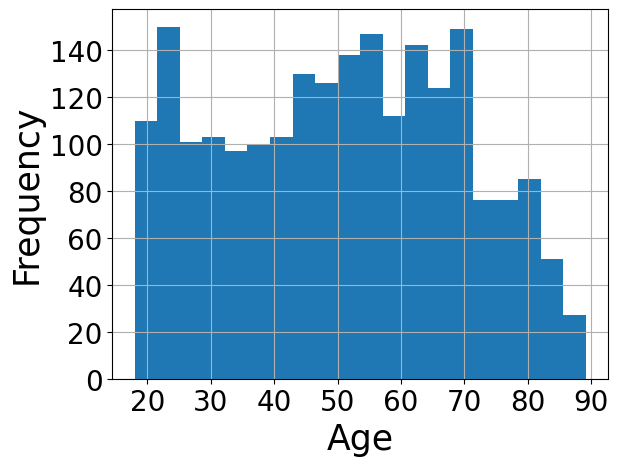}
   \caption{Age distribution of dataset used to train NeuroVNN.}
   \label{age_dist}
\end{figure}

We used the cortical thickness features derived from T1-weighted (T1w) MRI images collected from 3.0 Tesla MRI scanners for several publicly available datasets constituted by $2147$ healthy individuals who were 18 years or older. These datasets include: (a) Cambridge Centre for Ageing and Neuroscience (CamCAN) dataset~\cite{shafto2014cambridge}; (b) Dallas Lifespan Brain Study (DLBS) \href{https://fcon\textunderscore 1000.projects.nitrc.org/indi/retro/dlbs.html}{https://fcon\textunderscore 1000.projects.nitrc.org/indi/retro/dlbs.html}; (c) IXI dataset (\href{https://brain-development.org/ixi-dataset/}{https://brain-development.org/ixi-dataset/}; and enhanced Nathan Kline Institute-Rockland Sample (eNKI)~\cite{nooner2012nki}. The demographics of these datasets are summarized in Table~\ref{table_demo}. Figure~\ref{age_dist} provides the age distribution of the complete dataset. Although the combined dataset used to train NeuroVNN is of limited size, it uniformly represents individuals from different age groups in the age range 18-70, with the frequency tapering off for age group $> 70$.  The T1w MRI images in these datasets were pre-processed via the open-source CAT12 pipeline~\cite{gaser2022cat} to yield cortical thickness measures for each individual. In particular, for every individual, we obtained cortical thickness measures curated according to 100 parcellations, 200 parcellations, and 400 parcellations versions of Schaefer's brain atlas~\cite{schaefer2018local}, Desikan-Killiany (DK)~\cite{desikan2006automated} atlas consisting of $68$ parcellations, and DKT~\cite{destrieux2010automatic} atlas consisting of $148$ parcellations. NeuroVNN was trained on the cortical thickness dataset from population in Table~\ref{table_demo} that was curated according to 100 parcellations version of Schaefer's brain atlas.
\begin{table}[h]
\setlength{\tabcolsep}{3.5pt}
\caption{Demographics for datasets used for training NeuroVNN.}

\centering
\renewcommand{\arraystretch}{1}
\begin{threeparttable}
{\begin{tabular}{|c| c| c| c|c|}
\hline
\multirow{2}{*}{Dataset} & \multirow{2}{*}{n} & \multirow{2}{*}{Sex (m/f)} & \multicolumn{2}{c|}{Age}\\ 
\cline{4-5}
& & & range & mean $\pm$ s.d. \\
\hline
 CamCAN & 652 & 322/330 & 18.5-88.92 & 54.77 $\!\pm\!$ 18.6 \\
\hline
 DLBS & 315 & 117/198 & 20.57 -89.11& 54.62 $\pm$20.09 \\
 \hline
 IXI & 182& 88/94 &20.16-81.94 & 47.44$\pm$16.7 \\
 \hline
 eNKI & 998& 347/650\tnote{*} &18-85 &47.35$\pm$17.71 \\
\hline
 {\bf Total} & 2147 & 874/1272\tnote{*} &18-89.11 &50.68$\pm$18.62 \\
\hline
\end{tabular}}
\begin{tablenotes}\footnotesize
\item[*] Sex information missing for 1 individual.
\end{tablenotes}
\end{threeparttable}
\label{table_demo}
\end{table}

\begin{figure}[t]
  \centering
  \includegraphics[scale=0.35]{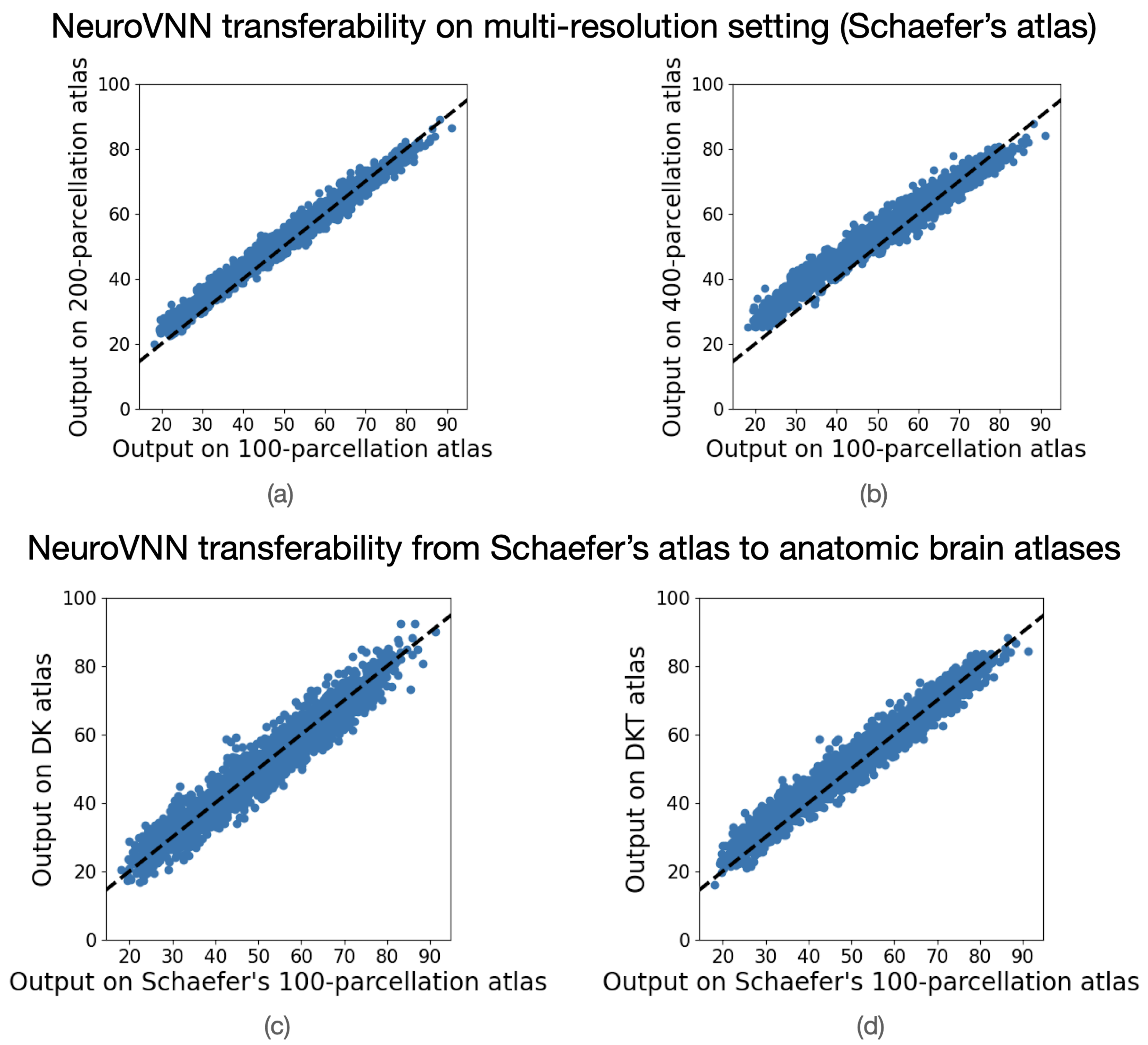}
   \caption{\textcolor{black}{One-sh}ot transferability of NeuroVNN for chronological age prediction task. NeuroVNN was trained on dataset curated according to 100 parcellations version of Schaefer's brain atlas and transferred to other versions of the dataset curated according to different brain atlases. }
   \label{vnn_transfer_training}
\end{figure}


\subsection{Architecture and Training}\label{training_details}
NeuroVNN was trained to predict the chronological age using $100$ cortical thickness features across the cortex for the population of healthy individuals, described in Table~\ref{table_demo}. The architecture consisted of $3$-layers, with $2$ filter taps each in the first and second layers, and $9$ filter taps in the third layer. The width of the network was set to $35$. The architecture parameters were selected based on the results of a hyperparameter search performed over $100$ trials using Optuna package~\cite{akiba2019optuna}. The training procedure was based on optimizing the mean-squared-error (MSE) loss function. The learning rate for the Adam optimizer was $0.01$. The model was trained on $90\%$ of the dataset and tested on the remaining $10\%$ to validate that it could infer information about chronological age on an unseen healthy population. This validation was necessary as brain age prediction relies on the ability of the model to exploit healthy ageing-related information in the cortical thickness features~\cite{sihag2023explainable}. The $90/10$ split of the dataset was determined randomly, and the constitution of the $10\%$ unseen data is provided in the supplementary material. The model operated on the anatomical covariance matrix of size $100\times 100$ estimated from the $100$ cortical thickness features of the $90\%$ split and normalized, such that its maximum eigenvalue was $1$. Also, the cortical thickness features were z-score normalized across the $90\%$ split and this normalization was applied to the $10\%$ validation cohort. Based on $10$ models trained on different permutations of the training set, NeuroVNN achieved an error performance of $9.24\pm0.59$ years with correlation between the ground truth and NeuroVNN estimates of chronological age being $0.79\pm 0.004$ on the test set. On the complete dataset, NeuroVNN achieved similar performance; the error was $9.16\pm 0.26$ years and the correlation between the ground truth and NeuroVNN estimates of the chronological age was $0.796\pm 0.0032$. We also observed that NeuroVNN's performance generalized to an independent dataset of $707$ adult individuals. Additional details are provided in Appendix~\ref{add_figs}.

\subsection{One-shot Transferability of NeuroVNN for Chronological Age Prediction Task}
NeuroVNN is a scale-free model and, hence, capable of processing a dataset of any arbitrary dimensionality. This feature is particularly relevant for data analyses in neuroimaging due to the prevalence of different brain atlases used to curate datasets. In terms of implementation, transferring a NeuroVNN from one cortical thickness dataset to another entails replacing the covariance matrix in the coVariance filters (i.e., the convolution operation). Because NeuroVNN is set up with a non-adaptive readout function, its architecture holds validity for dataset of any arbitrary dimensionality. 

Figure~\ref{vnn_transfer_training} demonstrates high similarity between the NeuroVNN outputs for the 100-dimensional dataset (that it was trained upon) and its outputs after being transferred to the 200, 400, 68, and 148 dimensional versions of the same dataset (Pearson's correlation $>0.97$ for all scenarios). The results for 200 and 400 parcellations in Fig\ref{vnn_transfer_training} correspond to different versions of Schaefer's 17-network atlas. 

One-shot transferability of NeuroVNN enables its transferability to datasets of different dimensionalities for the downstream tasks. This ability of NeuroVNN is remarkable, as it demonstrates that NeuroVNN can exploit the redundant information across datasets of different dimensionalities and is fundamental to the generalization of NeuroVNN. From a theoretical perspective, the quantitative transferability of VNN models has been studied previously across datasets whose covariance matrices are part of a converging sequence~\cite{sihag2023transferablility}.
\subsection{Anatomical Interpretability from NeuroVNN}
NeuroVNN architecture leverages a non-adaptive readout function to aggregate the outputs at its final layer to form the scalar chronological estimate. Thus, by projecting the final layer outputs of NeuroVNN on a brain surface, it is feasible, in principle, to quantify the `contribution' of each brain region to the chronological age estimate. This observation renders an \emph{anatomically interpretable} profile to the inference outcomes of NeuroVNN. Further, the outputs at the final layer of NeuroVNN are potential biomarkers as they can reflect changes in contributions for biologically plausible brain regions for a specific health condition. Specifically, as NeuroVNN is trained on healthy population, the variations in its outputs are expected to exploit the impacts on the underlying brain anatomy in various neurological contexts. This feature of VNNs was studied previously in~\cite{sihag2023explainable} to provide an anatomically interpretable perspective to brain age prediction in Alzheimer's disease (AD).

In this paper, we leverage the one-shot transferability property of NeuroVNN to demonstrate the generalizability of its anatomical interepretability to datasets of different dimensionalities. Specifically, as a foundation model, NeuroVNN has the ability to exploit biologically relevant features across datasets of different dimensionalities and provide consistent anatomical profiles to $\Delta$-Age derived after fine-tuning.

\subsection{Fine-tuning for Brain Age Prediction}
The statistical evidence on the training set thus far demonstrates that NeuroVNN estimates using the cortical thickness data correlate significantly with chronological age in healthy population, with consistent results even across datasets of dimensionalities different from that of the training set. Thus, in the presence of accelerated ageing (i.e., when an individual exhibits biological traits characteristic of someone older than their chronological age), it is expected that the final layer outputs of NeuroVNN will be elevated accordingly to reflect their increased contributions to the chronological age estimate formed by NeuroVNN. However, the chronological age estimates formed by NeuroVNN are far from accurate, as indicated by the mean error of $>9$ years and correlation of $~0.8$ on the training set in Section~\ref{training_details}. The lack of accuracy does not mitigate the ability of NeuroVNN to provide a meaningful brain age estimate, as the output of the NeuroVNN can be corrected or fine-tuned such that a clinician or the end user can detect accelerated biological ageing with respect to the chronological age of an individual. 

In general, the procedure for fine-tuning the chronological age estimates relies on a linear regression model. When NeuroVNN does not fit perfectly to the chronological age of healthy population, its residuals (i.e., the difference between the ground truth and estimates formed by NeuroVNN) are expected to be correlated with chronological age. Hence, the first step in the fine-tuning procedure involves learning a linear regression model from the healthy population that fits the residuals derived from NeuroVNN with the ground truth, i.e., the chronological age. For chronological age $y$ and the estimate $\hat y$
formed by NeuroVNN, the linear regression model is learned as
\begin{align}\label{s1}
     \hat y - y = \alpha y + \beta\;,
\end{align}
for scalar coefficients $\alpha$ and $\beta$. 
Next, the brain age estimate $\hat y_{\sf B}$ is derived as 
\begin{align}\label{s2}
   \hat y_{\sf B} = \hat y - (\alpha y + \beta)\;.
\end{align}
The difference between $\hat y_{\sf B}$ and $y$ is the brain age gap metric, i.e., $\Delta$-Age. For an individual of chronological age $y$, the brain age gap $\Delta$-Age is 
\begin{align}
     \Delta\text{-Age}\triangleq \hat y_{\sf B} - y\;,
\end{align}
where $\hat y_{\sf B}$ is evaluated from $\hat y$ and $y$ according~\eqref{s1} and~\eqref{s2}. The linear regression model in~\eqref{s1} is typically learnt only for healthy population and applied to other cohorts of interest in the dataset. The fine-tuning steps in~\eqref{s1} and~\eqref{s2} are formally referred to as age-bias correction steps in the existing literature on brain age prediction~\cite{de2020commentary}. 

From~\eqref{s1} and~\eqref{s2}, it can readily be concluded that the differences in brain age or $\Delta$-Age estimates for any two individuals with the same chronological age $y$ correlates with the difference in the corresponding NeuroVNN estimates. However, the evaluation of $\hat y_{\sf B}$ is contingent upon the data used to estimate coefficients $\alpha$ and $\beta$. Variations in the data used for estimating $\alpha$ and $\beta$ can naturally lead to variations in $\hat y_{\sf B}$ as well~\cite{more2023brain}, thus, rendering the objective of achieving uniqueness in brain age or $\Delta$-Age prediction an open problem. However, the statistical properties of the contributors to $\hat y$ are expected to reflect the statistical observations made by the analyses of $\hat y_{\sf B}$ or $\Delta$-Age. Specifically, if $\Delta$-Age was correlated with some biological variable or elevated for a health condition, we expected this effect to be reflected in similar analyses of certain outputs at the final layer of NeuroVNN. Mapping such outputs on the brain surface, in turn, can add to the anatomic interpretability and biological plausibility of $\Delta$-Age~\cite{sihag2023explainable}. 

Here, we focused on the qualitative aspect of brain age estimates formed by NeuroVNN in different contexts. Specifically, we investigate (i) whether the brain age derived from NeuroVNN that was trained only on the healthy population is biologically meaningful in different neurological contexts; and (ii) whether the biological plausibility of findings transfer to datasets of different dimensionalities. For this purpose, we used the pre-trained NeuroVNN model for brain age study across datasets of cognitively healthy populations and a population with AD. In Fig.~\ref{vnn_transfer_training}, it is apparent that NeuroVNN outputs are highly correlated but not identical across datasets associated with different brain atlases. Hence, the fine-tuning procedure was determined uniquely for each target dataset. 

\section{Results}\label{main_res}
Our experiments illuminate various practical aspects of NeuroVNN as a foundation model for brain age prediction. In particular, we focus on two datasets in the main paper: (i) the first dataset pertains to healthy individuals identified as at-risk for AD; a population which is one of the primary targets for machine learning-derived brain age predictions in clinical settings~\cite{baecker2021machine}; and (ii) the second dataset pertains to AD, in which the healthy individuals are $70$ years or older (which is an under-represented group in training dataset for NeuroVNN in Table~\ref{table_demo}).
\subsection{At-risk Cognitively Healthy Population}\label{preventad_exps}
\begin{figure}[t]
  \centering
  \includegraphics[scale=0.35]{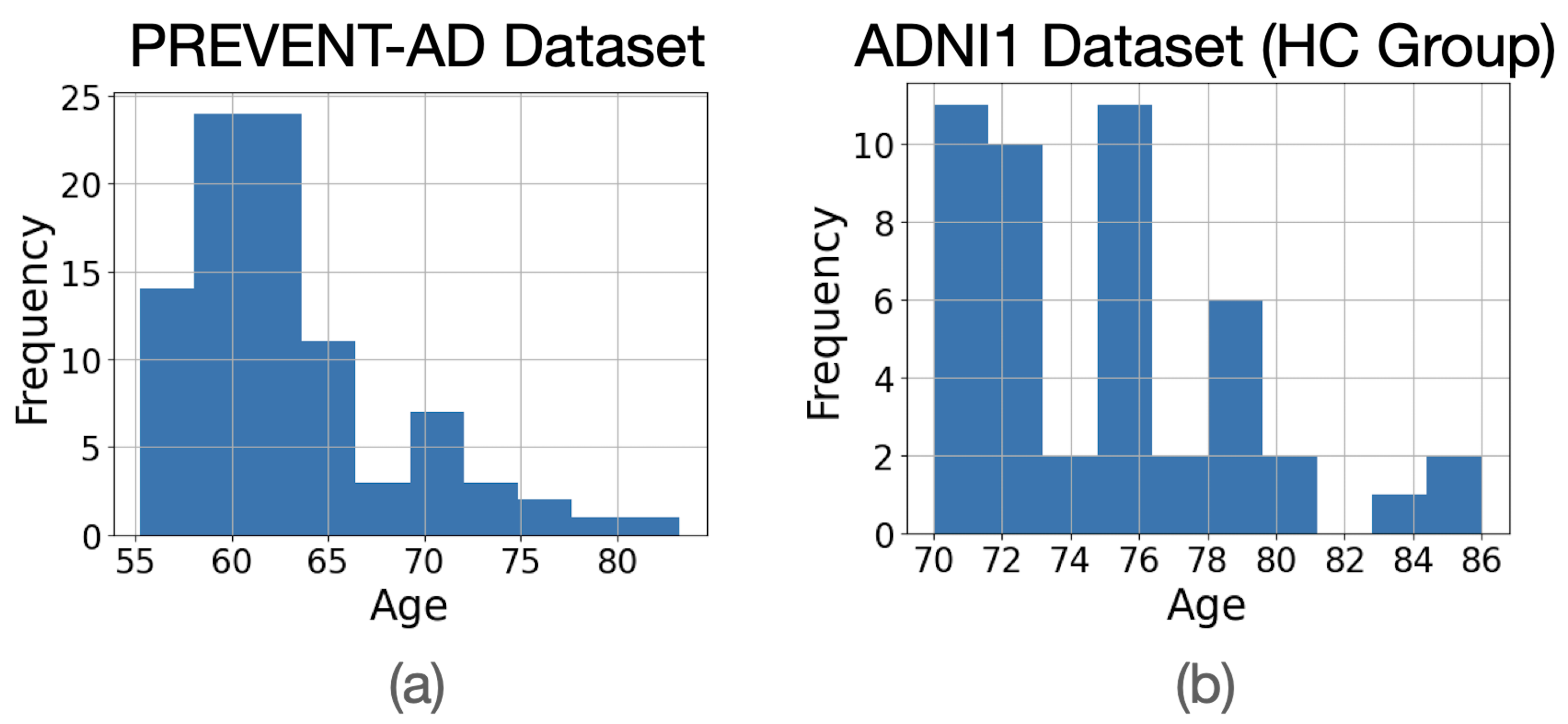}
   \caption{Age distributions in PREVENT-AD dataset and HC group of ADNI-1 dataset.}
   \label{prevent_ad_age}
\end{figure}

{\bf PREVENT-AD Dataset.} This dataset is a part of the PResymptomatic EValuation of Experimental or Novel Treatments for AD (PREVENT-AD) cohort~\cite{tremblay2021open} and consists of $90$ individuals (age = $62.95\pm 5.57$ years, $61$ females). The age distribution in this dataset is shown in Fig.~\ref{prevent_ad_age}. The individuals in this dataset were cognitively healthy and identified to be at-risk to developing AD due to parental or multiple-sibling history of AD or related dementias. Such factors put these individuals at an elevated risk of developing AD~\cite{huang2004apoe}. For each individual, we used the T1w MRI images from their baseline scan to derive cortical thickness metrics curated according to different brain atlases by the same pre-processing pipeline as described in Section~\ref{vnn_intro}. For all individuals, Alzheimer's progression score (APS) was available at the baseline visit. APS is a multimodal composite score calculated from multimodal imaging, neurosensory, cognitive, and cerebrospinal fluid markers, with increasing score indicating higher advancement of disease in the pre-symptomatic stage~\cite{leoutsakos2016alzheimer}.

\begin{figure}[H]
  \centering
  \includegraphics[scale=0.33]{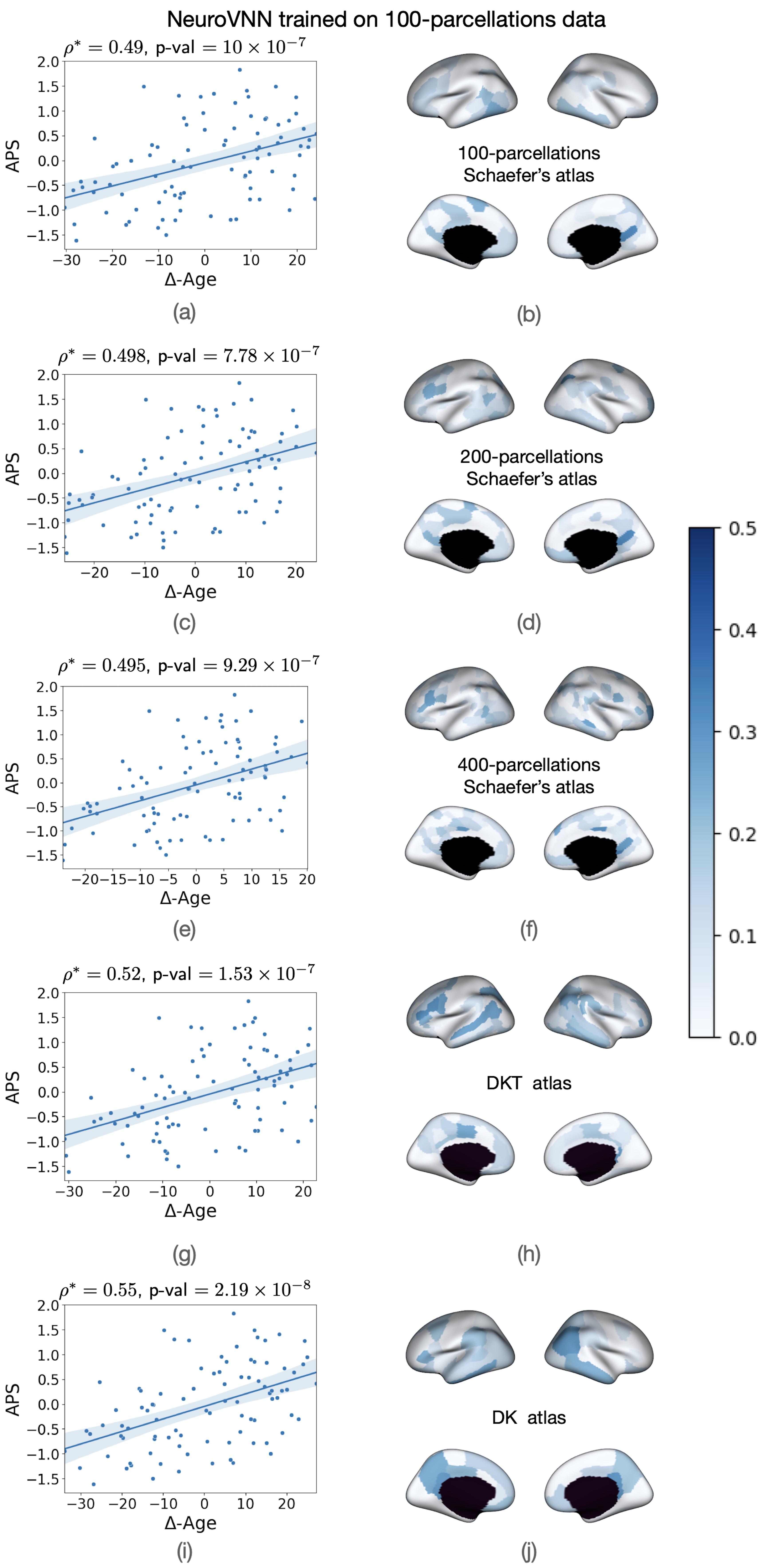}
   \caption{\textcolor{black}{Results on PREVENT-AD dataset obtained by NeuroVNN (trained on dataset with $100$ cortical thickness features). The figures in the right column plot the magnitude of positive correlation between APS and the associated entry at the final layer of NeuroVNN on a brain region, thus, providing an anatomic signature to the findings in the left column.}}
   \label{prevent_ad_1}
\end{figure}
\newcommand{\squeezeup}{\vspace{-2.5mm}}

\noindent
{\bf Experiment setup.} The age groups in PREVENT-AD dataset in Fig.~\ref{prevent_ad_age} are well-represented in general in the training dataset that NeuroVNN was trained upon (Fig.~\ref{age_dist}). We used NeuroVNN with the anatomical covariance matrix from the training data in Table~\ref{table_demo}. The biological plausibility of brain age estimates was investigated by evaluating correlation between the outputs at the final layer of NeuroVNN and APS as well as $\Delta$-Age estimate and APS. We hypothesized that if $\Delta$-Age derived using NeuroVNN was biologically plausible, it would be correlated with APS as higher $\Delta$-Age and higher APS are expected to be indicators of worsening impacts of neurodegeneration. 

\noindent
{\bf Observations.} For cortical thickness data curated according to $100$-parcellation Schaefer's atlas, $\Delta$-Age was significantly correlated with APS ($\rho = 0.43$, p-val = $2.9\times10^{-5}$; see Fig.~\ref{prevent_ad_1}a). Also, after correction for age and sex, the partial correlation between $\Delta$-Age and APS was significant (correlation after correction for age and sex, $\rho^{\ast}=0.49$, p-value $=10^{-6}$). Furthermore, Fig.~\ref{prevent_ad_1}b projects the correlations between the outputs at the final layer of NeuroVNN and APS. The brain regions in darker contrast in Fig.~\ref{prevent_ad_1}b provide the anatomical profile for the observed findings in Fig.~\ref{prevent_ad_1}a. The results in Fig.~\ref{prevent_ad_1}a-b provide the baseline to compare against in subsequent experiments pertaining to transferability of NeuroVNN.

Next, we tested whether the findings derived by NeuroVNN when transferred to cortical thickness datasets curated according to other brain atlases were consistent with those in Fig.~\ref{prevent_ad_1}a-b. When NeuroVNN was transferred to the versions of PREVENT-AD dataset that had been curated with Schaefer's 200 or 400-parcellations atlases, we observed results consistent with those in Fig.~\ref{prevent_ad_1}a-b in terms of correlation with APS as well as the anatomical profile associated with correlations between APS and outputs at the final layer of NeuroVNN (Fig.~\ref{prevent_ad_1}c-d and Fig.~\ref{prevent_ad_1}e-f). Notably, when NeuroVNN was transferred to datasets curated according to DK or DKT atlases, the partial correlation between $\Delta$-Age and APS was marginally higher ($0.52$ for DKT atlas and $0.55$ for DK atlas; see Fig.~\ref{prevent_ad_1}g and Fig.~\ref{prevent_ad_1}i). This marginal increase was accompanied by a more prominent involvement of regions in the temporal lobe in the anatomic profiles associated with the results in Fig.~\ref{prevent_ad_1}g and Fig.~\ref{prevent_ad_1}i. Atrophy in regions in the medial temporal lobe is characteristic of AD pathology with potential applications in clinical diagnosis of AD~\cite{duara2008medial}. Hence, the increase in correlation between APS and $\Delta$-Age in Fig.~\ref{prevent_ad_1}g and Fig.~\ref{prevent_ad_1}i relative to Fig.~\ref{prevent_ad_1}a, Fig.~\ref{prevent_ad_1}c, and Fig.~\ref{prevent_ad_1}e could be explained by the associated anatomic profiles derived by leveraging the outputs at the final layer of NeuroVNN. 


\subsection{Alzheimer's Disease}\label{adni1_res}
{\bf ADNI-1 Dataset.} This dataset is a standardized dataset provided as a part of Alzheimer's Disease Neuroimaging Initiative (ADNI) study(see~\cite{wyman2013standardization}) and consists of 47 healthy controls (HC; age = $75.06\pm 3.93$ years, 29 females), 71 individuals with mild cognitive impairment (MCI; age = $74.03\pm 8.12$ years, 26 females), and 33 individuals with dementia (AD; age = $75.08\pm 8.07$ years, 22 females). The age distribution of the individuals in the HC group is provided in Fig.~\ref{prevent_ad_age}b. For each individual, we used the 3.0T MRI images to derive cortical thickness metrics curated according to different brain atlases by the same pre-processing pipeline as described in Section~\ref{vnn_intro}.

\noindent
{\bf Experiment setup.} We hypothesized the $\Delta$-Age for AD and MCI groups to be elevated as compared to the HC group due to accelerated biological aging in AD~\cite{habes2016advanced,jove2014metabolomics}. The effectiveness of VNN-based brain age prediction framework in similar setting has been demonstrated previously in~\cite{sihag2023explainable}. This dataset is discussed here primarily because the age distribution of the healthy population in this dataset is skewed towards $70$ years or older, while the training dataset used for training NeuroVNN was relatively sparse in this age group as compared to other age groups (Fig.~\ref{age_dist}). Results on the broader ADNI database are included in Appendix~\ref{adni_full}. Since our objective was to detect accelerated biological ageing in MCI or AD groups, it was of interest to investigate whether NeuroVNN could provide meaningful $\Delta$-Age estimates despite the relatively sparse representation of older individuals in the dataset in Table~\ref{table_demo}. To this end, we first compared $\Delta$-Age estimates in the dataset of cortical thickness features curated according to Schaefer's $100$-parcellations brain atlas for NeuroVNN that used the covariance matrix from the dataset in Table~\ref{table_demo} and NeuroVNN that used the covariance matrix from the healthy population in ADNI-1 dataset (normalized such that the maximum eigenvalue was $1$). Note that in the latter scenario, the NeuroVNN was not re-trained and only the anatomical covariance matrix was replaced with that estimated from HC group in ADNI-1 dataset.

We also investigated the anatomic profile for elevated $\Delta$-Age in the combined AD and MCI cohort with respect to the HC cohort. For this purpose, the outputs at the final layer of NeuroVNN were compared for the combined AD and MCI cohort and HC cohort, and the outputs that were elevated in the combined AD and MCI cohorts with respect to those for the HC cohort were plotted on the brain surface.   Further, similar to the experiments in Section~\ref{preventad_exps}, whether the observations regarding $\Delta$-Age in  AD or MCI cohorts could be generalized across different datasets.

\noindent
{\bf Observations.} NeuroVNN with covariance matrix from the dataset in Table~\ref{table_demo} yielded elevated mean $\Delta$-Age in AD cohort (mean $\Delta$-Age = 6.9 years) with respect to HC cohort (mean $\Delta$-Age = 0 years), with $\Delta$-Age for MCI cohort between them (mean $\Delta$-Age = 3.6 years). However, ANCOVA with age and sex as covariates yielded no significant group differences among the $\Delta$-Age distributions for the three groups (p-value = $0.113$). On the other hand, NeuroVNN operating on the anatomical covariance matrix estimated from the HC group yielded larger $\Delta$-Age for AD cohort (mean $\Delta$-Age = 7.42 years) and MCI cohort (mean $\Delta$-Age = 5.3 years) relative to that for HC cohort (mean $\Delta$-Age = 0). Importantly, the group difference among the $\Delta$-Age for three cohorts was statistically significant (ANCOVA with age and sex as covariates: p-value = $0.001$). Thus, NeuroVNN was better equipped to detect accelerated aging in AD when operating on the covariance matrix from HC group, as compared to when it operated on the covariance matrix from the training dataset in Table~\ref{table_demo}. This observation could potentially be explained by the covariance matrix from the HC cohort in ADNI-1 dataset being better representative of the statistical properties relevant to $\Delta$-Age for this dataset as compared to the anatomical covariance matrix of the training dataset in Table~\ref{table_demo}, which may be biased towards the age group of individuals younger than $70$ years. Similar observations also extend to the wider ADNI database (see Appendix~\ref{adni_full}). Henceforth, we discuss the results obtained by NeuroVNN that operated on the covariance matrix derived from the HC cohort for the experiments on ADNI dataset.

\begin{figure}[H]
  \centering
  \includegraphics[scale=0.3]{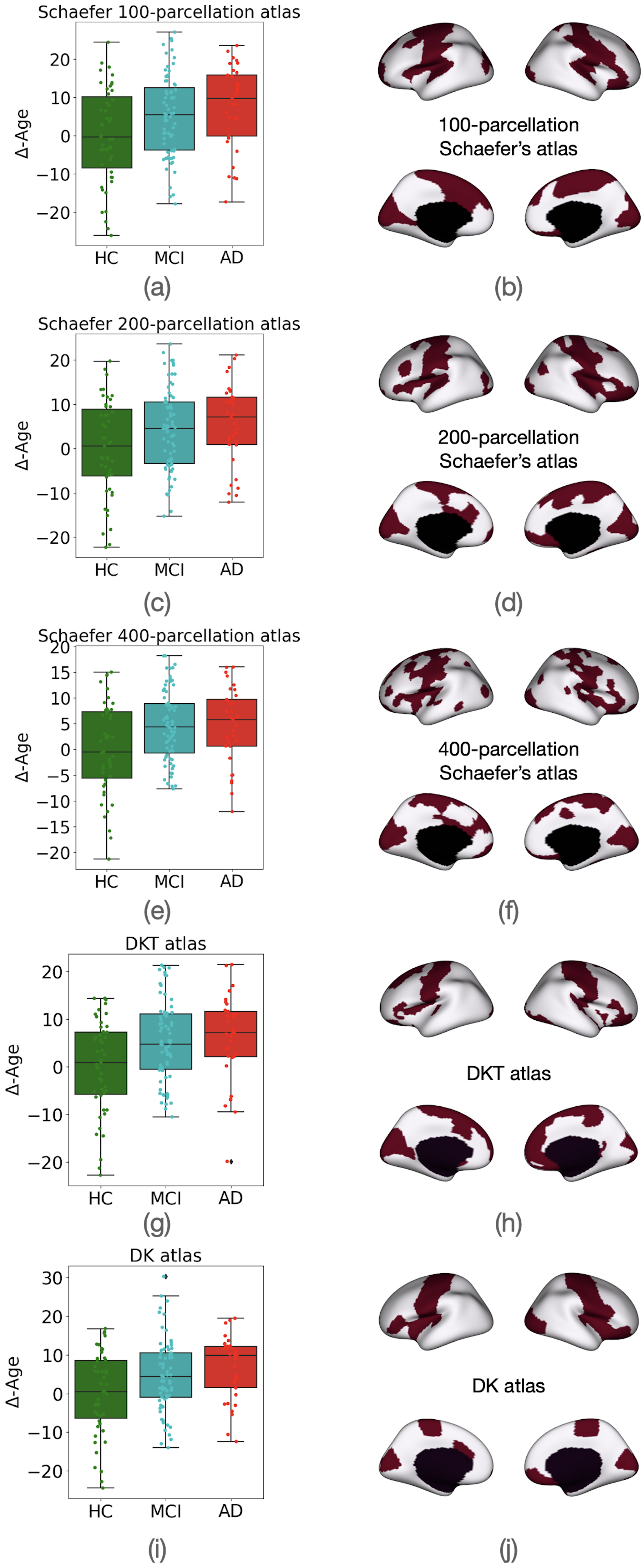}
   \caption{Results on ADNI-1 dataset obtained by NeuroVNN (trained on dataset with $100$ cortical thickness features). The figures in the right column highlight the brain regions for which the mean of the associated entry at the final layer of NeuroVNN was elevated in AD+MCI cohort relative to HC cohort.}
   \label{adni_1}
\end{figure}

Figure~\ref{adni_1}a plots the distributions of $\Delta$-Age for HC, MCI, and AD cohorts derived by NeuroVNN from the cortical thickness dataset curated according to 100-parcellations version of Schaefer's atlas; and Fig.~\ref{adni_1}b illustrates the anatomic regions associated with elevated $\Delta$-Age in MCI and AD cohorts with respect to the HC cohort. The brain regions isolated as contributors to elevated $\Delta$-Age in AD and MCI cohorts in Fig.~\ref{adni_1}b were consistent with the results reported in~\cite{sihag2023explainable} for the setting when VNN model operated on the anatomical covariance matrix estimated from the HC cohort. Subsequently, the results in Fig.~\ref{adni_1}a-b form the baseline to compare to when NeuroVNN was transferred to datasets curated according to other brain atlases. Figures~\ref{adni_1}c-j demonstrate the results derived using NeuroVNN on datasets curated according to different versions of Schaefer's atlas, DKT, and DK atlases. Notably, the anatomic profiles in Fig.~\ref{adni_1}d, f, h, and j associated with elevated $\Delta$-Age in AD and MCI cohorts were highly consistent with the results in Fig.~\ref{adni_1}b. Hence, NeuroVNN was able to transfer successfully to datasets with different dimensionalities to yield qualitatively consistent results for elevated $\Delta$-Age in AD and its associated anatomical profiles.

\section{Discussion}\label{discussion}
We have proposed NeuroVNN as a paradigm to construct foundation model for brain age prediction using neuroimaging data. NeuroVNN was trained on healthy population and shown to demonstrate a generalist characteristic for extracting $\Delta$-Age across various contexts. Specifically, NeuroVNN can provide meaningful biomarkers of neurodegeneration for various datasets that exhibited heterogeneity in terms of population characteristics and dimensionalities. Notably, the brain age estimates derived using VNN correlated with other known biomarkers of neurodegeneration in healthy populations that were at an elevated risk of dementia due to familial history and other biological factors. These findings suggest that NeuroVNN can add utility to tracking presymptomatic individuals and understanding mechanisms of neurodegeneration before clinical onset of the disease. From a data analyses perspective, NeuroVNN was able to extract $\Delta$-Age with consistent group-level statistical effects and associated anatomic signatures across datasets curated according to different brain atlases that partitioned the cortex into distinct number of regions. Thus, NeuroVNN is capable of exploiting the information relevant to $\Delta$-Age from neuroimaging datasets in a scale-free fashion, which can potentially advance computational neuroscience research in a new direction. 

Although NeuroVNN was trained on a dataset of limited size, the training dataset was representative of all age groups across the adult lifespan. It is of immediate interest to conduct a large-scale study with longitudinal analyses to evaluate the applicability of NeuroVNNs in addressing wider biological questions. Moreover, expanding the scope of experiments to a wider family of brain atlases can help gauge the limits of the transferability characteristic of NeuroVNN. 

We note that although $\Delta$-Age estimates exhibited consistent statistical effects at the group level after NeuroVNN was transferred across datasets, the $\Delta$-Age estimates were not identical at the individual level after transferring of NeuroVNN. In the absence of a ground truth, it is infeasible to comment on the optimality of the absolute value of $\Delta$-Age. However, the associated anatomical profiles with $\Delta$-Age in Fig.~\ref{prevent_ad_1} and Fig.~\ref{adni_1} were consistent, suggesting that NeuroVNN successfully exploited the underlying mechanism driving the statistical properties of $\Delta$-Age across datasets of different dimensionalities. Existing studies have reported variations in brain age estimates enabled by various methodological choices as a potential concern for practical applicability of brain age~\cite{more2023brain}. In principle, the anatomic interpretability offered by NeuroVNN can mitigate such concerns. However, if achieving identical $\Delta$-Age estimates across different datasets is the goal, the inclusion of domain-adaptation~\cite{yang2022data} and harmonization strategies~\cite{orlhac2022guide} with NeuroVNN could be explored. 

NeuroVNN has a generalist characteristic, that could be exploited further for biomarker discovery using a variety of downstream tasks (such as classification, prediction, clustering etc.) in future work. Traditionally, biomarker discovery has relied on statistical comparisons of different features between the cohorts representing adverse health conditions and the cohort representing the healthy control group. Our results for NeuroVNN have demonstrated that VNNs have the ability to provide a machine learning perspective to this conventional analyses by deploying models trained solely on healthy population and testing on cohorts with adverse health conditions.

\bibliographystyle{IEEEtran}
\bibliography{VNN_ICML}

\begin{thebibliography}{10}
\providecommand{\url}[1]{#1}
\csname url@samestyle\endcsname
\providecommand{\newblock}{\relax}
\providecommand{\bibinfo}[2]{#2}
\providecommand{\BIBentrySTDinterwordspacing}{\spaceskip=0pt\relax}
\providecommand{\BIBentryALTinterwordstretchfactor}{4}
\providecommand{\BIBentryALTinterwordspacing}{\spaceskip=\fontdimen2\font plus
\BIBentryALTinterwordstretchfactor\fontdimen3\font minus
  \fontdimen4\font\relax}
\providecommand{\BIBforeignlanguage}[2]{{%
\expandafter\ifx\csname l@#1\endcsname\relax
\typeout{** WARNING: IEEEtran.bst: No hyphenation pattern has been}%
\typeout{** loaded for the language `#1'. Using the pattern for}%
\typeout{** the default language instead.}%
\else
\language=\csname l@#1\endcsname
\fi
#2}}
\providecommand{\BIBdecl}{\relax}
\BIBdecl

\bibitem{lopez2013hallmarks}
C.~L{\'o}pez-Ot{\'\i}n \emph{et~al.}, ``The hallmarks of aging,'' \emph{Cell},
  vol. 153, no.~6, pp. 1194--1217, 2013.

\bibitem{fjell2010structural}
A.~M. Fjell and K.~B. Walhovd, ``Structural brain changes in aging: courses,
  causes and cognitive consequences,'' \emph{Reviews in the Neurosciences},
  vol.~21, no.~3, pp. 187--222, 2010.

\bibitem{cole2018brain}
J.~H. Cole, S.~J. Ritchie, M.~E. Bastin, V.~Hern{\'a}ndez,
  S.~Mu{\~n}oz~Maniega, N.~Royle, J.~Corley, A.~Pattie, S.~E. Harris, Q.~Zhang
  \emph{et~al.}, ``Brain age predicts mortality,'' \emph{Molecular Psychiatry},
  vol.~23, no.~5, pp. 1385--1392, 2018.

\bibitem{world2015world}
\emph{World report on ageing and health}.\hskip 1em plus 0.5em minus
  0.4em\relax World Health Organization, 2015.

\bibitem{johnson2009longitudinal}
D.~K. Johnson, M.~Storandt, J.~C. Morris, and J.~E. Galvin, ``Longitudinal
  study of the transition from healthy aging to {A}lzheimer disease,''
  \emph{Archives of neurology}, vol.~66, no.~10, pp. 1254--1259, 2009.

\bibitem{ferrucci2020measuring}
L.~Ferrucci, M.~Gonzalez-Freire, E.~Fabbri, E.~Simonsick, T.~Tanaka, Z.~Moore,
  S.~Salimi, F.~Sierra, and R.~de~Cabo, ``Measuring biological aging in humans:
  A quest,'' \emph{Aging cell}, vol.~19, no.~2, p. e13080, 2020.

\bibitem{cole2017predicting}
J.~H. Cole and K.~Franke, ``Predicting age using neuroimaging: Innovative brain
  ageing biomarkers,'' \emph{Trends in Neurosciences}, vol.~40, no.~12, pp.
  681--690, 2017.

\bibitem{baecker2021machine}
L.~Baecker, R.~Garcia-Dias, S.~Vieira, C.~Scarpazza, and A.~Mechelli, ``Machine
  learning for brain age prediction: Introduction to methods and clinical
  applications,'' \emph{EBioMedicine}, vol.~72, p. 103600, 2021.

\bibitem{baecker2021brain}
L.~Baecker, J.~Dafflon, P.~F. Da~Costa, R.~Garcia-Dias, S.~Vieira,
  C.~Scarpazza, V.~D. Calhoun, J.~R. Sato, A.~Mechelli, and W.~H. Pinaya,
  ``Brain age prediction: A comparison between machine learning models using
  region-and voxel-based morphometric data,'' \emph{Human Brain Mapping},
  vol.~42, no.~8, pp. 2332--2346, 2021.

\bibitem{beheshti2021predicting}
I.~Beheshti, M.~Ganaie, V.~Paliwal, A.~Rastogi, I.~Razzak, and M.~Tanveer,
  ``Predicting brain age using machine learning algorithms: A comprehensive
  evaluation,'' \emph{IEEE Journal of Biomedical and Health Informatics},
  vol.~26, no.~4, pp. 1432--1440, 2021.

\bibitem{pina2022structural}
O.~Pina, I.~Cumplido-Mayoral, R.~Cacciaglia, J.~M. Gonz{\'a}lez-de
  Ech{\'a}varri, J.~D. Gispert, and V.~Vilaplana, ``Structural networks for
  brain age prediction,'' in \emph{International Conference on Medical Imaging
  with Deep Learning}.\hskip 1em plus 0.5em minus 0.4em\relax PMLR, 2022, pp.
  944--960.

\bibitem{franke2019ten}
K.~Franke and C.~Gaser, ``Ten years of brainage as a neuroimaging biomarker of
  brain aging: What insights have we gained?'' \emph{Frontiers in neurology},
  p. 789, 2019.

\bibitem{franke2012longitudinal}
------, ``Longitudinal changes in individual brainage in healthy aging, mild
  cognitive impairment, and {A}lzheimer’s disease.'' \emph{GeroPsych: The
  Journal of Gerontopsychology and Geriatric Psychiatry}, vol.~25, no.~4, p.
  235, 2012.

\bibitem{sihag2023explainable}
\BIBentryALTinterwordspacing
S.~Sihag, G.~Mateos, C.~McMillan, and A.~Ribeiro, ``Explainable brain age
  prediction using covariance neural networks,'' in \emph{Thirty-seventh
  Conference on Neural Information Processing Systems}, 2023. [Online].
  Available: \url{https://openreview.net/forum?id=cAhJF87GN0}
\BIBentrySTDinterwordspacing

\bibitem{bommasani2021opportunities}
R.~Bommasani, D.~A. Hudson, E.~Adeli, R.~Altman, S.~Arora, S.~von Arx, M.~S.
  Bernstein, J.~Bohg, A.~Bosselut, E.~Brunskill \emph{et~al.}, ``On the
  opportunities and risks of foundation models,'' \emph{arXiv preprint
  arXiv:2108.07258}, 2021.

\bibitem{lee2022deep}
J.~Lee, B.~J. Burkett, H.-K. Min, M.~L. Senjem, E.~S. Lundt, H.~Botha,
  J.~Graff-Radford, L.~R. Barnard, J.~L. Gunter, C.~G. Schwarz \emph{et~al.},
  ``Deep learning-based brain age prediction in normal aging and dementia,''
  \emph{Nature Aging}, vol.~2, no.~5, pp. 412--424, 2022.

\bibitem{karim2021aging}
H.~T. Karim, M.~Ly, G.~Yu, R.~Krafty, D.~L. Tudorascu, H.~J. Aizenstein, and
  C.~Andreescu, ``Aging faster: {W}orry and rumination in late life are
  associated with greater brain age,'' \emph{Neurobiology of Aging}, vol. 101,
  pp. 13--21, 2021.

\bibitem{liem2017predicting}
F.~Liem, G.~Varoquaux, J.~Kynast, F.~Beyer, S.~K. Masouleh, J.~M. Huntenburg,
  L.~Lampe, M.~Rahim, A.~Abraham, R.~C. Craddock \emph{et~al.}, ``Predicting
  brain-age from multimodal imaging data captures cognitive impairment,''
  \emph{Neuroimage}, vol. 148, pp. 179--188, 2017.

\bibitem{ronan2016obesity}
L.~Ronan, A.~F. Alexander-Bloch, K.~Wagstyl, S.~Farooqi, C.~Brayne, L.~K.
  Tyler, P.~C. Fletcher \emph{et~al.}, ``Obesity associated with increased
  brain age from midlife,'' \emph{Neurobiology of aging}, vol.~47, pp. 63--70,
  2016.

\bibitem{mareckova2020maternal}
K.~Mareckova, R.~Marecek, L.~Andryskova, M.~Brazdil, and Y.~S. Nikolova,
  ``Maternal depressive symptoms during pregnancy and brain age in young adult
  offspring: findings from a prenatal birth cohort,'' \emph{Cerebral Cortex},
  vol.~30, no.~7, pp. 3991--3999, 2020.

\bibitem{westlye2012effects}
L.~T. Westlye, I.~Reinvang, H.~Rootwelt, and T.~Espeseth, ``Effects of apoe on
  brain white matter microstructure in healthy adults,'' \emph{Neurology},
  vol.~79, no.~19, pp. 1961--1969, 2012.

\bibitem{thomas2020dealing}
R.~M. Thomas, W.~Bruin, P.~Zhutovsky, and G.~van Wingen, ``Dealing with missing
  data, small sample sizes, and heterogeneity in machine learning studies of
  brain disorders,'' in \emph{Machine learning}.\hskip 1em plus 0.5em minus
  0.4em\relax Elsevier, 2020, pp. 249--266.

\bibitem{lawrence2021standardizing}
R.~M. Lawrence, E.~W. Bridgeford, P.~E. Myers, G.~C. Arvapalli, S.~C.
  Ramachandran, D.~A. Pisner, P.~F. Frank, A.~D. Lemmer, A.~Nikolaidis, and
  J.~T. Vogelstein, ``Standardizing human brain parcellations,''
  \emph{Scientific data}, vol.~8, no.~1, p.~78, 2021.

\bibitem{yang2022data}
Y.~Yang, Y.~Zhu, H.~Cui, X.~Kan, L.~He, Y.~Guo, and C.~Yang, ``Data-efficient
  brain connectome analysis via multi-task meta-learning,'' \emph{arXiv
  preprint arXiv:2206.04486}, 2022.

\bibitem{sihag2022covariance}
S.~Sihag, G.~Mateos, C.~McMillan, and A.~Ribeiro, ``{coVariance} neural
  networks,'' in \emph{Proc. Conference on Neural Information Processing
  Systems}, Nov. 2022.

\bibitem{sihag2023transferablility}
S.~Sihag, G.~Mateos, C.~T. McMillan, and A.~Ribeiro, ``Transferablility of
  covariance neural networks and application to interpretable brain age
  prediction using anatomical features,'' \emph{arXiv preprint
  arXiv:2305.01807}, 2023.

\bibitem{lombardi2021explainable}
A.~Lombardi, D.~Diacono, N.~Amoroso, A.~Monaco, J.~M.~R. Tavares, R.~Bellotti,
  and S.~Tangaro, ``Explainable deep learning for personalized age prediction
  with brain morphology,'' \emph{Frontiers in neuroscience}, vol.~15, p. 578,
  2021.

\bibitem{yin2023anatomically}
C.~Yin, P.~Imms, M.~Cheng, A.~Amgalan, N.~F. Chowdhury, R.~J. Massett, N.~N.
  Chaudhari, X.~Chen, P.~M. Thompson, P.~Bogdan \emph{et~al.}, ``Anatomically
  interpretable deep learning of brain age captures domain-specific cognitive
  impairment,'' \emph{Proceedings of the National Academy of Sciences}, vol.
  120, no.~2, p. e2214634120, 2023.

\bibitem{brown2020language}
T.~Brown, B.~Mann, N.~Ryder, M.~Subbiah, J.~D. Kaplan, P.~Dhariwal,
  A.~Neelakantan, P.~Shyam, G.~Sastry, A.~Askell \emph{et~al.}, ``Language
  models are few-shot learners,'' \emph{Advances in neural information
  processing systems}, vol.~33, pp. 1877--1901, 2020.

\bibitem{yuan2021florence}
L.~Yuan, D.~Chen, Y.-L. Chen, N.~Codella, X.~Dai, J.~Gao, H.~Hu, X.~Huang,
  B.~Li, C.~Li \emph{et~al.}, ``Florence: A new foundation model for computer
  vision,'' \emph{arXiv preprint arXiv:2111.11432}, 2021.

\bibitem{yu2022coca}
J.~Yu, Z.~Wang, V.~Vasudevan, L.~Yeung, M.~Seyedhosseini, and Y.~Wu, ``Coca:
  Contrastive captioners are image-text foundation models,'' \emph{arXiv
  preprint arXiv:2205.01917}, 2022.

\bibitem{wu2023visual}
C.~Wu, S.~Yin, W.~Qi, X.~Wang, Z.~Tang, and N.~Duan, ``Visual chatgpt: Talking,
  drawing and editing with visual foundation models,'' \emph{arXiv preprint
  arXiv:2303.04671}, 2023.

\bibitem{zhou2023comprehensive}
C.~Zhou, Q.~Li, C.~Li, J.~Yu, Y.~Liu, G.~Wang, K.~Zhang, C.~Ji, Q.~Yan, L.~He
  \emph{et~al.}, ``A comprehensive survey on pretrained foundation models: A
  history from bert to chatgpt,'' \emph{arXiv preprint arXiv:2302.09419}, 2023.

\bibitem{kirillov2023segment}
A.~Kirillov, E.~Mintun, N.~Ravi, H.~Mao, C.~Rolland, L.~Gustafson, T.~Xiao,
  S.~Whitehead, A.~C. Berg, W.-Y. Lo \emph{et~al.}, ``Segment anything,''
  \emph{arXiv preprint arXiv:2304.02643}, 2023.

\bibitem{moor2023foundation}
M.~Moor, O.~Banerjee, Z.~S.~H. Abad, H.~M. Krumholz, J.~Leskovec, E.~J. Topol,
  and P.~Rajpurkar, ``Foundation models for generalist medical artificial
  intelligence,'' \emph{Nature}, vol. 616, no. 7956, pp. 259--265, 2023.

\bibitem{wojcik2022foundation}
M.~A. W{\'o}jcik, ``Foundation models in healthcare: Opportunities, biases and
  regulatory prospects in europe,'' in \emph{International Conference on
  Electronic Government and the Information Systems Perspective}.\hskip 1em
  plus 0.5em minus 0.4em\relax Springer, 2022, pp. 32--46.

\bibitem{steinberg2021language}
E.~Steinberg, K.~Jung, J.~A. Fries, C.~K. Corbin, S.~R. Pfohl, and N.~H. Shah,
  ``Language models are an effective representation learning technique for
  electronic health record data,'' \emph{Journal of biomedical informatics},
  vol. 113, p. 103637, 2021.

\bibitem{sihag2022predicting}
S.~Sihag, G.~Mateos, C.~McMillan, and A.~Ribeiro, ``Predicting brain age using
  transferable {coVariance} neural networks,'' in \emph{Proc. IEEE
  International Conference on Acoustics, Speech, and Signal Processing}, Jun.
  2023.

\bibitem{more2023brain}
S.~More, G.~Antonopoulos, F.~Hoffstaedter, J.~Caspers, S.~B. Eickhoff, K.~R.
  Patil, {A}lzheimer's Disease Neuroimaging~Initiative \emph{et~al.},
  ``Brain-age prediction: A systematic comparison of machine learning
  workflows,'' \emph{NeuroImage}, vol. 270, p. 119947, 2023.

\bibitem{butler2021pitfalls}
E.~R. Butler, A.~Chen, R.~Ramadan, T.~T. Le, K.~Ruparel, T.~M. Moore, T.~D.
  Satterthwaite, F.~Zhang, H.~Shou, R.~C. Gur \emph{et~al.}, ``Pitfalls in
  brain age analyses,'' Wiley Online Library, Tech. Rep., 2021.

\bibitem{gama2020graphs}
F.~Gama, E.~Isufi, G.~Leus, and A.~Ribeiro, ``Graphs, convolutions, and neural
  networks: From graph filters to graph neural networks,'' \emph{IEEE Signal
  Processing Magazine}, vol.~37, no.~6, pp. 128--138, 2020.

\bibitem{gama2020stability}
F.~Gama, J.~Bruna, and A.~Ribeiro, ``Stability properties of graph neural
  networks,'' \emph{IEEE Transactions on Signal Processing}, vol.~68, pp.
  5680--5695, 2020.

\bibitem{shafto2014cambridge}
M.~A. Shafto, L.~K. Tyler, M.~Dixon, J.~R. Taylor, J.~B. Rowe, R.~Cusack, A.~J.
  Calder, W.~D. Marslen-Wilson, J.~Duncan, T.~Dalgleish \emph{et~al.}, ``The
  cambridge centre for ageing and neuroscience (cam-can) study protocol: a
  cross-sectional, lifespan, multidisciplinary examination of healthy cognitive
  ageing,'' \emph{BMC neurology}, vol.~14, pp. 1--25, 2014.

\bibitem{nooner2012nki}
K.~B. Nooner, S.~J. Colcombe, R.~H. Tobe, M.~Mennes, M.~M. Benedict, A.~L.
  Moreno, L.~J. Panek, S.~Brown, S.~T. Zavitz, Q.~Li \emph{et~al.}, ``The
  nki-rockland sample: a model for accelerating the pace of discovery science
  in psychiatry,'' \emph{Frontiers in neuroscience}, vol.~6, p. 152, 2012.

\bibitem{gaser2022cat}
C.~Gaser, R.~Dahnke, P.~M. Thompson, F.~Kurth, E.~Luders, and {A}lzheimer’s
  Disease Neuroimaging~Initiative, ``{CAT}--a computational anatomy toolbox for
  the analysis of structural mri data,'' \emph{biorxiv}, pp. 2022--06, 2022.

\bibitem{schaefer2018local}
A.~Schaefer, R.~Kong, E.~M. Gordon, T.~O. Laumann, X.-N. Zuo, A.~J. Holmes,
  S.~B. Eickhoff, and B.~T. Yeo, ``Local-global parcellation of the human
  cerebral cortex from intrinsic functional connectivity {{MRI}},''
  \emph{Cerebral Cortex}, vol.~28, no.~9, pp. 3095--3114, 2018.

\bibitem{desikan2006automated}
R.~S. Desikan, F.~S{\'e}gonne, B.~Fischl, B.~T. Quinn, B.~C. Dickerson,
  D.~Blacker, R.~L. Buckner, A.~M. Dale, R.~P. Maguire, B.~T. Hyman
  \emph{et~al.}, ``An automated labeling system for subdividing the human
  cerebral cortex on {{MRI}} scans into gyral based regions of interest,''
  \emph{Neuroimage}, vol.~31, no.~3, pp. 968--980, 2006.

\bibitem{destrieux2010automatic}
C.~Destrieux, B.~Fischl, A.~Dale, and E.~Halgren, ``Automatic parcellation of
  human cortical gyri and sulci using standard anatomical nomenclature,''
  \emph{Neuroimage}, vol.~53, no.~1, pp. 1--15, 2010.

\bibitem{akiba2019optuna}
T.~Akiba, S.~Sano, T.~Yanase, T.~Ohta, and M.~Koyama, ``Optuna: A
  next-generation hyperparameter optimization framework,'' in \emph{Proceedings
  of the 25th ACM SIGKDD International Conference on Knowledge Discovery \&
  Data Mining}, 2019, pp. 2623--2631.

\bibitem{de2020commentary}
A.-M.~G. de~Lange and J.~H. Cole, ``Commentary: Correction procedures in
  brain-age prediction,'' \emph{NeuroImage: Clinical}, vol.~26, 2020.

\bibitem{tremblay2021open}
J.~Tremblay-Mercier, C.~Madjar, S.~Das, A.~P. Binette, S.~O. Dyke,
  P.~{\'E}tienne, M.-E. Lafaille-Magnan, J.~Remz, P.~Bellec, D.~L. Collins
  \emph{et~al.}, ``Open science datasets from {PREVENT-AD}, a longitudinal
  cohort of pre-symptomatic {A}lzheimer’s disease,'' \emph{NeuroImage:
  Clinical}, vol.~31, p. 102733, 2021.

\bibitem{huang2004apoe}
W.~Huang, C.~Qiu, E.~von Strauss, B.~Winblad, and L.~Fratiglioni, ``Apoe
  genotype, family history of dementia, and {A}lzheimer disease risk: a 6-year
  follow-up study,'' \emph{Archives of neurology}, vol.~61, no.~12, pp.
  1930--1934, 2004.

\bibitem{leoutsakos2016alzheimer}
J.-M. Leoutsakos, A.~Gross, R.~Jones, M.~Albert, and J.~Breitner,
  ``‘{A}lzheimer’s progression score’: development of a biomarker summary
  outcome for ad prevention trials,'' \emph{The journal of prevention of
  {A}lzheimer's disease}, vol.~3, no.~4, p. 229, 2016.

\bibitem{duara2008medial}
R.~Duara, D.~Loewenstein, E.~Potter, J.~Appel, M.~Greig, R.~Urs, Q.~Shen,
  A.~Raj, B.~Small, W.~Barker \emph{et~al.}, ``Medial temporal lobe atrophy on
  mri scans and the diagnosis of {A}lzheimer disease,'' \emph{Neurology},
  vol.~71, no.~24, pp. 1986--1992, 2008.

\bibitem{wyman2013standardization}
B.~T. Wyman, D.~J. Harvey, K.~Crawford, M.~A. Bernstein, O.~Carmichael, P.~E.
  Cole, P.~K. Crane, C.~DeCarli, N.~C. Fox, J.~L. Gunter \emph{et~al.},
  ``Standardization of analysis sets for reporting results from adni mri
  data,'' \emph{{A}lzheimer's \& Dementia}, vol.~9, no.~3, pp. 332--337, 2013.

\bibitem{habes2016advanced}
M.~Habes, D.~Janowitz, G.~Erus, J.~Toledo, S.~Resnick, J.~Doshi, S.~Van~der
  Auwera, K.~Wittfeld, K.~Hegenscheid, N.~Hosten \emph{et~al.}, ``Advanced
  brain aging: Relationship with epidemiologic and genetic risk factors, and
  overlap with {A}lzheimer disease atrophy patterns,'' \emph{Translational
  Psychiatry}, vol.~6, no.~4, pp. e775--e775, 2016.

\bibitem{jove2014metabolomics}
M.~Jov{\'e}, M.~Portero-Ot{\'\i}n, A.~Naud{\'\i}, I.~Ferrer, and R.~Pamplona,
  ``Metabolomics of human brain aging and age-related neurodegenerative
  diseases,'' \emph{Journal of Neuropathology \& Experimental Neurology},
  vol.~73, no.~7, pp. 640--657, 2014.

\bibitem{orlhac2022guide}
F.~Orlhac, J.~J. Eertink, A.-S. Cottereau, J.~M. Zijlstra, C.~Thieblemont,
  M.~Meignan, R.~Boellaard, and I.~Buvat, ``A guide to combat harmonization of
  imaging biomarkers in multicenter studies,'' \emph{Journal of Nuclear
  Medicine}, vol.~63, no.~2, pp. 172--179, 2022.

\bibitem{taylor2017cambridge}
J.~R. Taylor, N.~Williams, R.~Cusack, T.~Auer, M.~A. Shafto, M.~Dixon, L.~K.
  Tyler, R.~N. Henson \emph{et~al.}, ``The cambridge centre for ageing and
  neuroscience (cam-can) data repository: Structural and functional mri, meg,
  and cognitive data from a cross-sectional adult lifespan sample,''
  \emph{neuroimage}, vol. 144, pp. 262--269, 2017.

\bibitem{harms2018extending}
M.~P. Harms, L.~H. Somerville, B.~M. Ances, J.~Andersson, D.~M. Barch,
  M.~Bastiani, S.~Y. Bookheimer, T.~B. Brown, R.~L. Buckner, G.~C. Burgess
  \emph{et~al.}, ``Extending the human connectome project across ages: Imaging
  protocols for the lifespan development and aging projects,''
  \emph{Neuroimage}, vol. 183, pp. 972--984, 2018.

\bibitem{roberts2018prevalence}
R.~O. Roberts, J.~A. Aakre, W.~K. Kremers, M.~Vassilaki, D.~S. Knopman, M.~M.
  Mielke, R.~Alhurani, Y.~E. Geda, M.~M. Machulda, P.~Coloma \emph{et~al.},
  ``Prevalence and outcomes of amyloid positivity among persons without
  dementia in a longitudinal, population-based setting,'' \emph{JAMA
  neurology}, vol.~75, no.~8, pp. 970--979, 2018.

\bibitem{sperling2020association}
R.~A. Sperling, M.~C. Donohue, R.~Raman, C.-K. Sun, R.~Yaari, K.~Holdridge,
  E.~Siemers, K.~A. Johnson, P.~S. Aisen, A.~S. Team \emph{et~al.},
  ``Association of factors with elevated amyloid burden in clinically normal
  older individuals,'' \emph{JAMA neurology}, vol.~77, no.~6, pp. 735--745,
  2020.

\bibitem{fagan2015does}
A.~M. Fagan, ``What does it mean to be ‘amyloid-positive’?'' \emph{Brain},
  vol. 138, no.~3, pp. 514--516, 2015.

\bibitem{hampel2021amyloid}
H.~Hampel, J.~Hardy, K.~Blennow, C.~Chen, G.~Perry, S.~H. Kim, V.~L.
  Villemagne, P.~Aisen, M.~Vendruscolo, T.~Iwatsubo \emph{et~al.}, ``The
  amyloid-$\beta$ pathway in {A}lzheimer’s disease,'' \emph{Molecular
  psychiatry}, vol.~26, no.~10, pp. 5481--5503, 2021.

\bibitem{simren2021diagnostic}
J.~Simr{\'e}n, A.~Leuzy, T.~K. Karikari, A.~Hye, A.~L. Benedet,
  J.~Lantero-Rodriguez, N.~Mattsson-Carlgren, M.~Sch{\"o}ll, P.~Mecocci,
  B.~Vellas \emph{et~al.}, ``The diagnostic and prognostic capabilities of
  plasma biomarkers in {A}lzheimer's disease,'' \emph{{A}lzheimer's \&
  Dementia}, vol.~17, no.~7, pp. 1145--1156, 2021.

\bibitem{benedet2019plasma}
A.~L. Benedet, N.~J. Ashton, T.~A. Pascoal, A.~Leuzy, S.~Mathotaarachchi, M.~S.
  Kang, J.~Therriault, M.~Savard, M.~Chamoun, M.~Sch{\"o}ll \emph{et~al.},
  ``Plasma neurofilament light associates with {A}lzheimer's disease metabolic
  decline in amyloid-positive individuals,'' \emph{{A}lzheimer's \& Dementia:
  Diagnosis, Assessment \& Disease Monitoring}, vol.~11, no.~1, pp. 679--689,
  2019.

\bibitem{braak1993staging}
H.~Braak, E.~Braak, and J.~Bohl, ``Staging of {A}lzheimer-related cortical
  destruction,'' \emph{European neurology}, vol.~33, no.~6, pp. 403--408, 1993.

\end{thebibliography}

\newpage
\appendix
\onecolumn

\section{Acknowledgements}
CamCAN dataset used in the preparation of this work were obtained from the CamCAN repository (available at \href{http://www.mrc-cbu.cam.ac.uk/datasets/camcan/}{http://www.mrc-cbu.cam.ac.uk/datasets/camcan/})\cite{taylor2017cambridge,shafto2014cambridge}. Dallas Lifespan Brain Study (DLBS) is available at International Neuroimaging Data-sharing Initiative (INDI). IXI dataset is available at \href{https://brain-development.org/ixi-dataset/}{https://brain-development.org/ixi-dataset/}. The eNKI dataset is available at \href{https://fcon\textunderscore 1000.projects.nitrc.org/indi/pro/eNKI\textunderscore RS\textunderscore TRT/FrontPage.html}{https://fcon\textunderscore 1000.projects.nitrc.org/indi/pro/eNKI\textunderscore RS\textunderscore TRT/FrontPage.html}. PREVENT-AD dataset used were obtained from the
Pre-symptomatic Evaluation of Novel or Experimental Treatments for Alzheimer's Disease (PREVENT-AD)
program~\cite{tremblay2021open}.The cortical thickness measures for HCP-A dataset were downloaded from the NIMH Data Archive. 

The datasets from the Alzheimer’s Disease
Neuroimaging Initiative (ADNI) database were obtained from ~\href{adni.loni.usc.edu}{adni.loni.usc.edu}. As such, the investigators
within the ADNI contributed to the design and implementation of ADNI and/or provided data
but did not participate in analysis or writing of this report. Data collection and sharing for ADNI was funded by the Alzheimer's Disease
Neuroimaging Initiative (ADNI) (National Institutes of Health Grant U19 AG024904) and
DOD ADNI (Department of Defense award number W81XWH-12-2-0012). ADNI is funded
by the National Institute on Aging, the National Institute of Biomedical Imaging and
Bioengineering, and through generous contributions from the following: AbbVie, Alzheimer’s
Association; Alzheimer’s Drug Discovery Foundation; Araclon Biotech; BioClinica, Inc.;
Biogen; Bristol-Myers Squibb Company; CereSpir, Inc.; Cogstate; Eisai Inc.; Elan
Pharmaceuticals, Inc.; Eli Lilly and Company; EuroImmun; F. Hoffmann-La Roche Ltd and
its affiliated company Genentech, Inc.; Fujirebio; GE Healthcare; IXICO Ltd.; Janssen
Alzheimer Immunotherapy Research and Development, LLC.; Johnson and Johnson
Pharmaceutical Research and Development LLC.; Lumosity; Lundbeck; Merck and Co., Inc.;
Meso Scale Diagnostics, LLC.; NeuroRx Research; Neurotrack Technologies; Novartis
Pharmaceuticals Corporation; Pfizer Inc.; Piramal Imaging; Servier; Takeda Pharmaceutical
Company; and Transition Therapeutics. The Canadian Institutes of Health Research is
providing funds to support ADNI clinical sites in Canada. Private sector contributions are
facilitated by the Foundation for the National Institutes of Health (www.fnih.org). The grantee
organization is the Northern California Institute for Research and Education, and the study is
coordinated by the Alzheimer’s Therapeutic Research Institute at the University of Southern
California. ADNI data are disseminated by the Laboratory for Neuro Imaging at the
University of Southern California.

\section{Additional Results and Experiment Details}\label{add_figs}
\begin{figure}[h]
  \centering
  \includegraphics[scale=0.3]{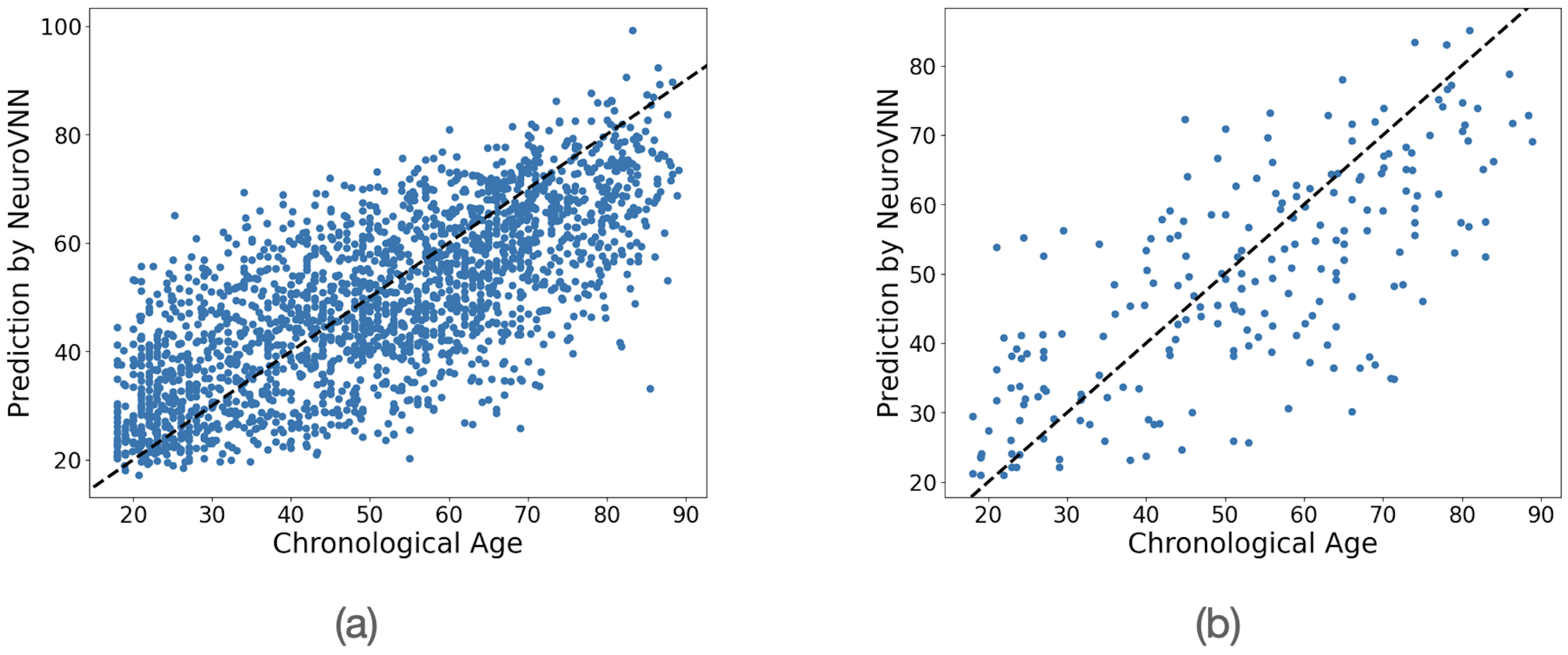}
   \caption{Plots of chronological age estimates formed by NeuroVNN versus the ground truth for the complete dataset in Table~\ref{table_demo} and the test set described in Section~\ref{training_details}, respectively. }
   \label{train_test_complete}
\end{figure}

Figure~\ref{train_test_complete}a and b plot the chronological age estimates formed by NeuroVNN versus the ground truth for the complete dataset in Table~\ref{table_demo} and the test set described in Section~\ref{training_details}, respectively. The test set consisted of 80 individuals from CamCAN dataset, 24 individuals from DLBS, 12 individuals from IXI, and 98 individuals from eNKI dataset. 

\begin{figure}[h]
  \centering
  \includegraphics[scale=0.3]{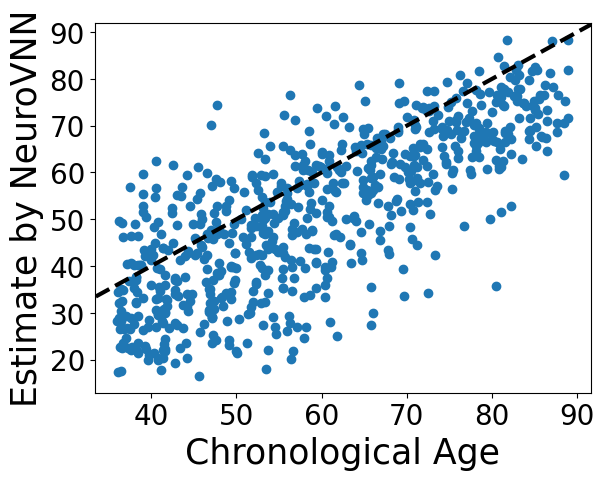}
   \caption{Plot of estimates formed by NeuroVNN versus chronological age in HCP-A dataset.}
   \label{hcp_result}
\end{figure}

{\bf External validation for NeuroVNN on HCP-A dataset for the chronological age prediction task.} HCP-A dataset is part of the Human Connectome Project~\cite{harms2018extending} and consists of $707$ healthy individuals within the age range of $36-89$ (mean age: $59.41$ years, standard deviation: $14.76$ years, $395$ females). $18$ individuals older than $89$ were excluded to match the age-range of individuals in this dataset with the dataset in Table~\ref{table_demo}. This dataset consisted of $68$ cortical thickness features curated according to DK atlas and was provided by the National Institute of Mental Health Data Archive (\href{https://nda.nih.gov/}{https://nda.nih.gov/}). Figure~\ref{hcp_result} plots the chronological age estimates obtained by NeuroVNN on this dataset versus the ground truth. Despite NeuroVNN being trained on a different brain atlas (100 parcellations version of Schaefer's brain atlas), its outputs were significantly correlated with chronological age in this dataset (Pearson's correlation = $0.77$). Further, NeuroVNN achieved the mean absolute error of $10.63$. This experiment demonstrated the ability of NeuroVNN to exploit age-related information on an independent dataset of dimensionality different from that of its training dataset. 

\section{Results on ADNI Dataset (DK atlas)}\label{adni_full}
\begin{figure}[h]
  \centering
  \includegraphics[scale=0.3]{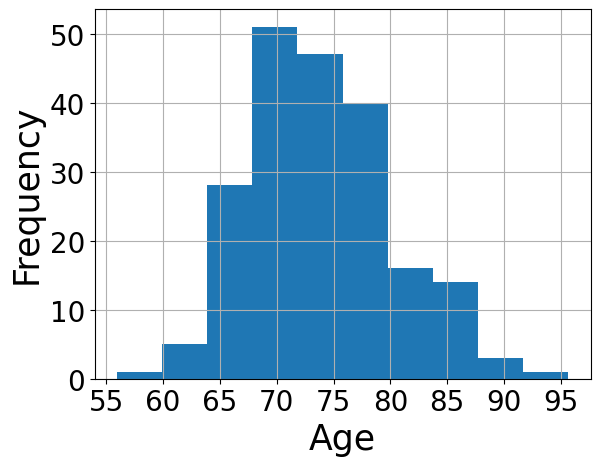}
   \caption{\textcolor{black}{Age distribution in HC group in ADNI dataset}.}
   \label{adni_age_dist}
\end{figure}

{\bf Dataset.} This dataset constituted of $206$ healthy individuals (HC), $372$ individuals with MCI, and $118$ individuals diagnosed with dementia (AD). The demographics of this dataset are summarized in Table~\ref{adni_demo}. The age distribution of the HC group is illustrated in Fig.~\ref{adni_age_dist}. For each individual, we used the $68$ cortical thickness features across the cortex. This dataset was downloaded from \href{https://adni.loni.usc.edu/}{https://adni.loni.usc.edu/}. The cortical thickness features available at \href{https://adni.loni.usc.edu/}{https://adni.loni.usc.edu/} were derived using Freesurfer 5.1 pipeline for T1w images from structural MRI scans and curated according to Desikan-Killiany brain atlas.

\begin{table}[h]
\setlength{\tabcolsep}{3.5pt}
\caption{Demographics for ADNI dataset.}

\centering
\renewcommand{\arraystretch}{1}
\begin{threeparttable}
{\begin{tabular}{|c| c| c| c|c|}
\hline
\multirow{2}{*}{Category} & \multirow{2}{*}{n} & \multirow{2}{*}{Sex (m/f)} & \multicolumn{2}{c|}{Age}\\ 
\cline{4-5}
& & & range & mean $\pm$ s.d. \\
\hline
HC & 206 & 96/106 & 55.95-95.62 & 73.87 $\pm$ 6.39\\
\hline
MCI & 372 & 212/160 & 55.49-92.11 & 72.25 $\pm$ 7.61\\
\hline
AD & 118 & 62/56 & 55.78-91.42 & 73.84 $\pm$ 7.56\\
\hline
\end{tabular}}
\end{threeparttable}
\label{adni_demo}
\end{table}

{\bf Experiment setup.} The experimental setup is similar to that for the ADNI-1 dataset in Section~\ref{adni1_res}. We hypothesized the AD and MCI groups to exhibit elevated $\Delta$-Age relative to HC group due to accelerated biological aging in AD. Similar to ADNI-1 dataset, the age distribution of HC group in the wider ADNI dataset is also skewed towards individuals in $70+$ age group. Hence, we investigated whether the anatomical covariance matrix from the training dataset in Table~\ref{table_demo} or the anatomical covariance matrix from the HC group in ADNI dataset (normalized such that the maximum eigenvalue was $1$) were a better choice for brain age prediction using NeuroVNN in this dataset.

\begin{figure}[h]
  \centering
  \includegraphics[scale=0.3]{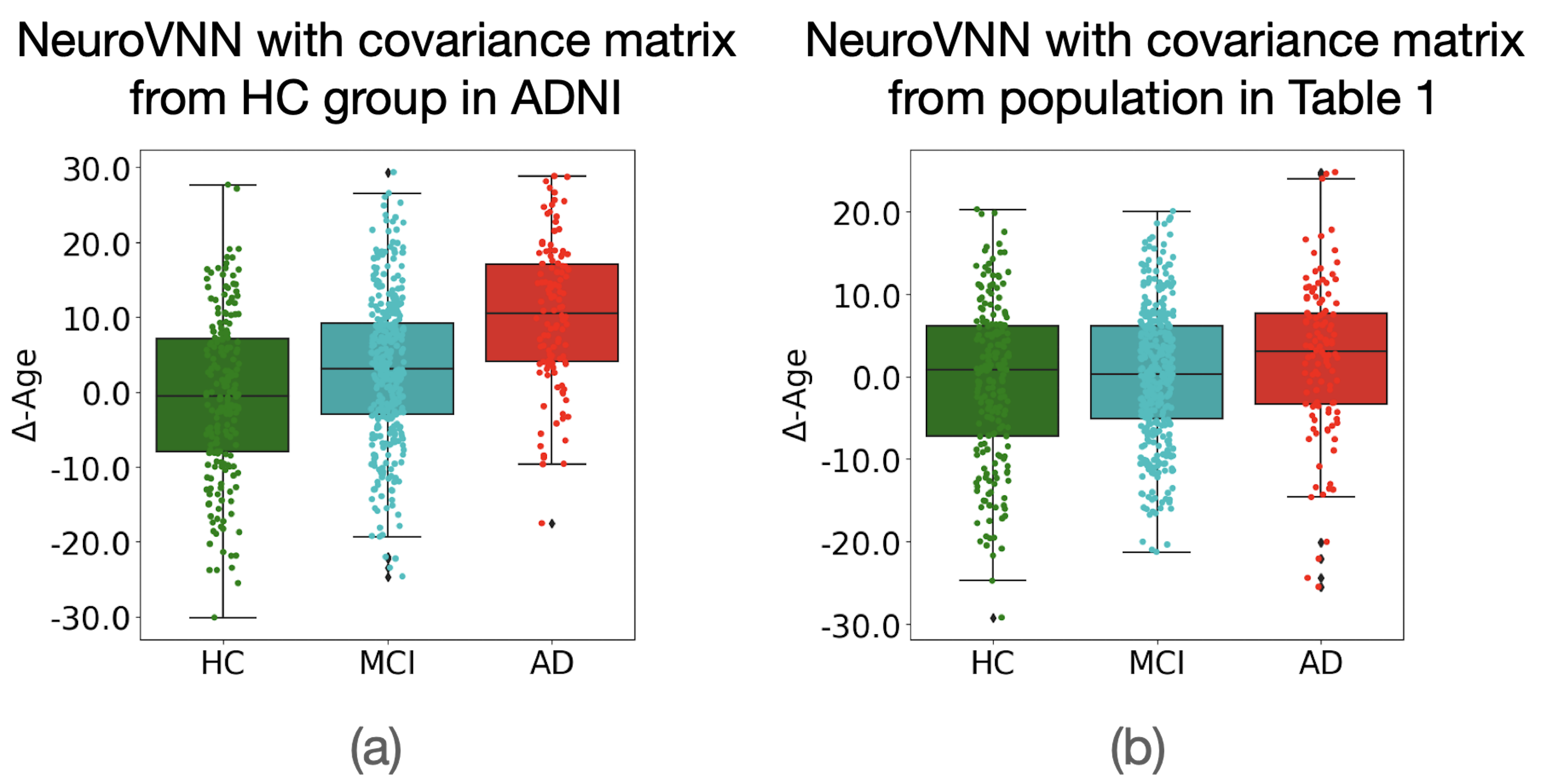}
   \caption{Comparison between $\Delta$-Age distributions derived from (a) NeuroVNN operating on covariance matrix from HC group in ADNI and (b) NeuroVNN operating on covariance matrix from the dataset in Table~\ref{table_demo}.}
   \label{neurovnn_adni_comp}
\end{figure}

\noindent
{\bf Observations.} Consistent with the observations in Section~\ref{adni_1}, NeuroVNN operating on the covariance matrix estimated from the cortical thickness features in the HC group in ADNI dataset provided more discriminative $\Delta$-Age in AD and MCI groups relative to HC group as compared to the results obtained using NeuroVNN operating on the covariance matrix estimated from the dataset in Table~\ref{table_demo}. Figure~\ref{neurovnn_adni_comp} plots the distributions of $\Delta$-Age in HC, MCI, and AD groups obtained from NeuroVNN operating on covariance matrix from HC group in ADNI and NeuroVNN operating on covariance matrix from the dataset. For Fig.~\ref{neurovnn_adni_comp}a, the means $\Delta$-Age for MCI and AD groups are $3.09$ and $10.33$ years, respectively. Further, the group differences among the $\Delta$-Age distributions for HC, MCI, and AD groups are significant (ANCOVA with age and sex as covariates: p-value $<10e-8$). On the other hand, the means of $\Delta$-Age for AD and MCI groups in Fig.~\ref{neurovnn_adni_comp}b are $2.21$ years and $0.22$ years, respectively. Thus, NeuroVNN was better equipped to detect accelerated aging in AD when operating on the covariance matrix from HC group. 

\section{$\Delta$-Age in Amyloid Positive Healthy Individuals in ADNI Dataset}\label{adni_nfl}
We further extended the experiments in Appendix~\ref{adni_full} to investigate $\Delta$-Age in the amyloid positive subgroup of the individuals in the HC group. There were $59$ amyloid positive individuals and $147$ amyloid negative individuals in the HC group of the ADNI dataset studied in Appendix~\ref{adni_full} (amyloid status available on \href{https://adni.loni.usc.edu}{https://adni.loni.usc.edu}).

\noindent
{\bf Background.} Existing clinical studies report that amyloid positive, cognitively healthy individuals are at relatively elevated risk of developing cognitive impairment~\cite{roberts2018prevalence, sperling2020association}. Moreover, amyloid $\beta$ deposition in the brain has been at the center of Alzheimer's disease research and clinical trials~\cite{fagan2015does,hampel2021amyloid}. Furthermore, plasma neurofilament light chain (NfL) is a promising biofluid biomarker with potential uses in screening, diagnosis, and monitoring of AD~\cite{simren2021diagnostic}, which may be associated with changes in brain anatomy in AD-specific brain regions for amyloid positive individuals~\cite{benedet2019plasma}. Thus, amyloid positive individuals in the HC group could be categorized as at-risk for cognitive impairment and biological progression towards AD.

\noindent
{\bf Experiment setup.} We aimed to check whether $\Delta$-Age predicted by NeuroVNN for the amyloid positive individuals correlated with plasma NfL, and whether such correlations had biologically plausible anatomic profile. For this purpose, we focused on $33$ amyloid positive individuals for whom the plasma NfL data collected within $30$ days of the MRI scan was available. 
\begin{figure}[ht]
  \centering
  \includegraphics[scale=0.3]{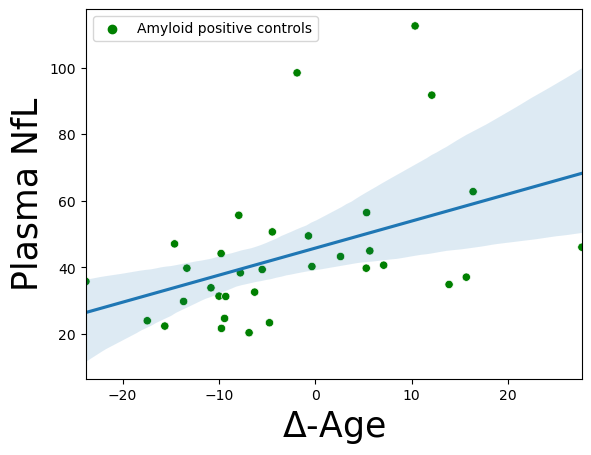}
   \caption{$\Delta$-Age correlated with plasma NfL in amyloid positive individuals in the HC group.}
   \label{neurovnn_nfl}
\end{figure}

\begin{figure}[ht]
  \centering
  \includegraphics[scale=0.3]{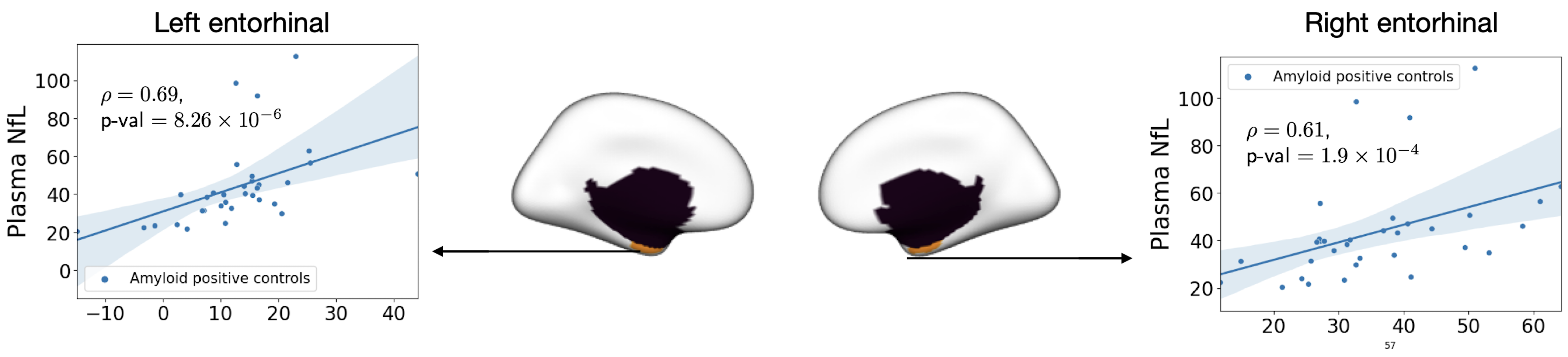}
   \caption{Regional contributors to chronological age estimates from left and right entorhinal regions correlated with plasma NfL in amyloid positive controls.}
   \label{neurovnn_nfl_entorhinal}
\end{figure}
{\bf Observations. } Our experiments showed that there was no significant difference in $\Delta$-Age between the amyloid positive and amyloid negative cohorts in the HC group. For the cohort of amyloid positive individuals with plasma NfL and MRI scans collected within 30 days of each other, we observed a positive correlation between $\Delta$-Age and plasma NfL (partial correlation after correction for age and sex: $0.43$, p-value = 0.015; Fig.~\ref{neurovnn_nfl}). Notably, the anatomic profile derived from the correlation analyses at the final layer of the NeuroVNN identified left entorhinal and right entorhinal brain regions as most significantly correlated with plasma NfL (Fig.~\ref{neurovnn_nfl_entorhinal}). This observation is significant because entorhinal region is one of the first regions impacted by AD~\cite{braak1993staging}. Thus, the results in Section~\ref{preventad_exps}, the results on ADNI dataset together provided evidence across two independent datasets that $\Delta$-Age derived using NeuroVNN could be reflective of impending neurodegeneration due to Alzheimer's disease.

\clearpage
\end{document}